\documentclass{article}

\usepackage[preprint]{neurips_2025}

\usepackage[utf8]{inputenc} 
\usepackage[T1]{fontenc}    
\usepackage{hyperref}       
\usepackage{url}            
\usepackage{booktabs}       
\usepackage{amsfonts}       
\usepackage{nicefrac}       
\usepackage{microtype}      
\usepackage{xcolor}         
\usepackage{amsmath}
\usepackage{multirow}
\usepackage{threeparttable}
\usepackage{graphicx}
\usepackage{amssymb}
\usepackage{bm}
\usepackage[ruled,linesnumbered]{algorithm2e}
\usepackage{placeins}
\usepackage{authblk}

\title{EEG-fused Digital Twin Brain for Autonomous Driving in Virtual Scenarios}

\author{}

\begin{document}

\maketitle
\vspace{-4.5em}

{
  Yubo Hou\textsuperscript{1},
  Zhengxin Zhang\textsuperscript{2},
  Ziyi Wang\textsuperscript{1},
  Wenlian Lu\textsuperscript{1,2$\ast$},
  Jianfeng Feng\textsuperscript{1,2$\ast$},
  Taiping Zeng\textsuperscript{2$\ast$}
\\
  \textit{1} School of Mathematical Sciences, Fudan University, Shanghai 200433, China 
\\
  \textit{2} Institute of Science and Technology for Brain-inspired Intelligence, Fudan University, Shanghai 
\\
  200433, China
\\
  $\ast$ Corresponding authors: 
  \\
  Taiping Zeng, Institute of Science and Technology for Brain-inspired Intelligence, Fudan University, No. 220 Handan Road, Shanghai, 200433, China. Email: zengtaiping@fudan.edu.cn \\
  Jianfeng Feng, Institute of Science and Technology for Brain-inspired Intelligence, Fudan University, No. 220 Handan Road, Shanghai, 200433, China. Email: jianfeng64@gmail.com \\
  Wenlian Lu, School of Mathematical Sciences, Fudan University, No. 220 Handan Road, Shanghai, 200433, China. Email: wenlian@fudan.edu.cn   
}

\vspace{2.5em}

\begin{abstract}
  Current methodologies typically integrate biophysical brain models with functional magnetic resonance imaging(fMRI) data - while offering millimeter-scale spatial resolution (0.5-2 $mm^{3}$ voxels), these approaches suffer from limited temporal resolution (>0.5 Hz) for tracking rapid neural dynamics during continuous tasks. Conversely, Electroencephalogram (EEG) provides millisecond-scale temporal precision ($\le 1$ ms sampling rate) for real-time guidance of continuous task execution, albeit constrained by low spatial resolution. To reconcile these complementary modalities, we present a generalizable Bayesian inference framework that integrates high-spatial-resolution structural MRI(sMRI) with high-temporal-resolution EEG to construct a biologically realistic digital twin brain(DTB) model. The framework establishes voxel-wise mappings between millisecond-scale EEG and sMRI-derived spiking networks, while demonstrating its translational potential through a brain-inspired autonomous driving simulation. Our EEG-DTB model achieves capabilities: (1) Biologically-plausible EEG signal generation (0.88 resting-state,0.60 task-state correlation), with simulated signals in task-state yielding steering predictions outperforming both chance and empirical signals (p<0.05); (2) Successful autonomous driving in the CARLA simulator using decoded steering angles. The proposed approach pioneers a new paradigm for studying sensorimotor integration and for mechanistic studies of perception-action cycles and the development of brain-inspired control systems.
  
\end{abstract}

\section{Introduction}

Simulating brain dynamics has become an indispensable approach for investigating the neural mechanisms underlying perceptual motion and decoding information flow along neural pathways. To achieve this, the construction of biologically plausible brain models requires: (1) selecting appropriate neuronal dynamics representations, and (2) effectively integrating multimodal neuroimaging data to constrain model parameters.

Current neuronal dynamic models primarily fall into two categories: biophysical models based on simulating dynamic behavior of neuronal membrane potential, such as spiking neuron models~\cite{stein1965theoretical, burkitt2006review} and Hodgkin-Huxley (HH) models~\cite{hodgkin1952quantitative}, and population-level models based on the mean field or random field method, such as Wilson-Cowan model~\cite{wilson1972excitatory}. While these neuronal models establish theoretical foundations for brain simulation, their biological fidelity requires integration with empirical neuroimaging data.

Recent advances in neuroimaging technologies, including functional magnetic resonance imaging (fMRI) ~\cite{friston2014nodes}, positron emission tomography (PET) ~\cite{Pogossian1975}, electroencephalogram (EEG) ~\cite{berger1931elektrenkephalogramm, michel2012towards} and magnetoencephalogram (MEG) ~\cite{boto2018moving}, provide complementary structural and functional data at multiple scales. However, integrating multimodal measurements with different spatiotemporal scales, such as MRI and EEG, to infer parameters in large-scale networks remains a challenge.

Data assimilation(DA) is a mathematical technology that fine-tunes model parameters based on observational data, enabling real-time simulation of system state. While widely applied in fields such as geosciences ~\cite{eibern1999four} and finance ~\cite{Lopes2011}, its adoption in computational neuroscience is still emerging ~\cite{politi2016comparing}. Recent studies have proposed a hierarchical data assimilation (HDA) framework that assimilates fMRI signals into spiking neuronal networks through estimating model hyperparameters~\cite{zhang2024framework}. Although fMRI offers superior spatial resolution(0.5-2 $mm^{3}$ voxels), it is limited by its temporal resolution(>0.5 Hz) to capture rapid dynamic oscillations.  

In contrast to fMRI, EEG offers millisecond temporal resolution, making it particularly suited for studying information encoding and transmission in neural circuits. Recent works have successfully simulated EEG signals based on neural mass models using DA approaches~\cite{escuain2018extracranial, yokoyama2023data, jansen1995electroencephalogram}. However, these methods face two fundamental limitations: (1) the inherently poor spatial resolution of EEG (typically centimeter-scale) ~\cite{Ramezani2025}, and (2) an inability to infer parameters at the synaptic level.

\begin{figure}
  \centering
  \includegraphics[width=5.5in]{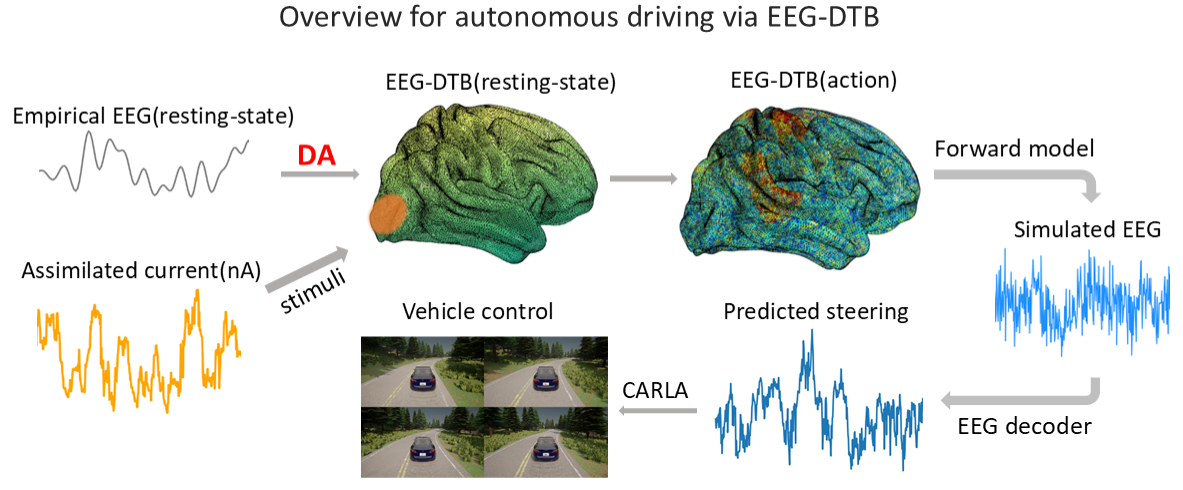}
  \caption{ {\bf Overview for autonomous driving via EEG-Digital twin brain(EEG-DTB).} The resting-state EEG-DTB parameters are established by assimilating empirical resting-state EEG data. Task-state activation is then induced through visual pathway current stimuli. The simulated post-synaptic currents generate task-state EEG signals via the MRI-constrained forward model, which are decoded into steering angles for vehicle control during curve navigation in the CARLA simulator.}
  \label{fig:fig}
\end{figure}

\begin{figure}
  \centering
  \includegraphics[width=5.5in]{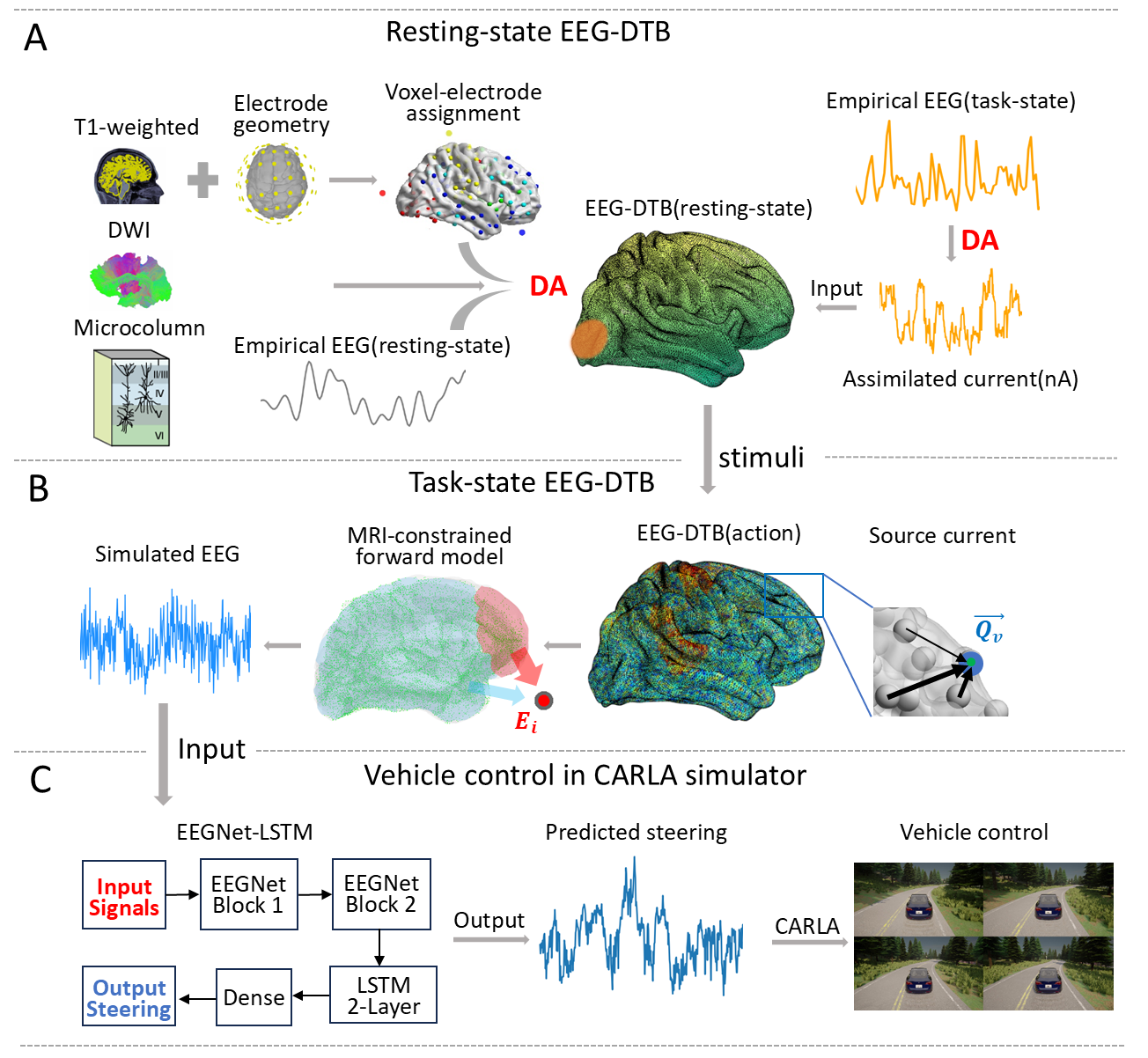}
  \caption{ {\bf Pipeline for EEG-Digital twin brain(EEG-DTB) construction and autonomous driving in virtual scenarios.}The pipeline comprises three phases: (A) Resting-state EEG-DTB. The cortico-subcortical model is built from multimodal MRI(including DWI and T1-weighted MRI) and microcolumn connection map, with voxel-electrode assignments derived from EEG forward modeling(based on T1w anatomy and electrode geometry) to reconcile the spatial resolution disparities between EEG and MRI. Synaptic conductances are determined through assimilating resting-state EEG signals into the cortico-subcortical model, generating resting-state EEG-DTB. (B) Task-state EEG-DTB. Visual-associated regions are stimulated with currents inferred from task-state EEG, generating simulated EEG signals via our MRI-constrained forward model, where the magnitudes and orientations of source currents are determined by the post-synaptic currents and voxel positions, respectively. (C) Vehicle control in the CARLA simulator. The simulated EEG signals are decoded via the EEGNet-LSTM model to predict steering angles, which are then executed in the CARLA simulator to control vehicle turning maneuvers.}
  \label{fig:fig1}
\end{figure}

\begin{figure}
  \centering
  \includegraphics[width=5in]{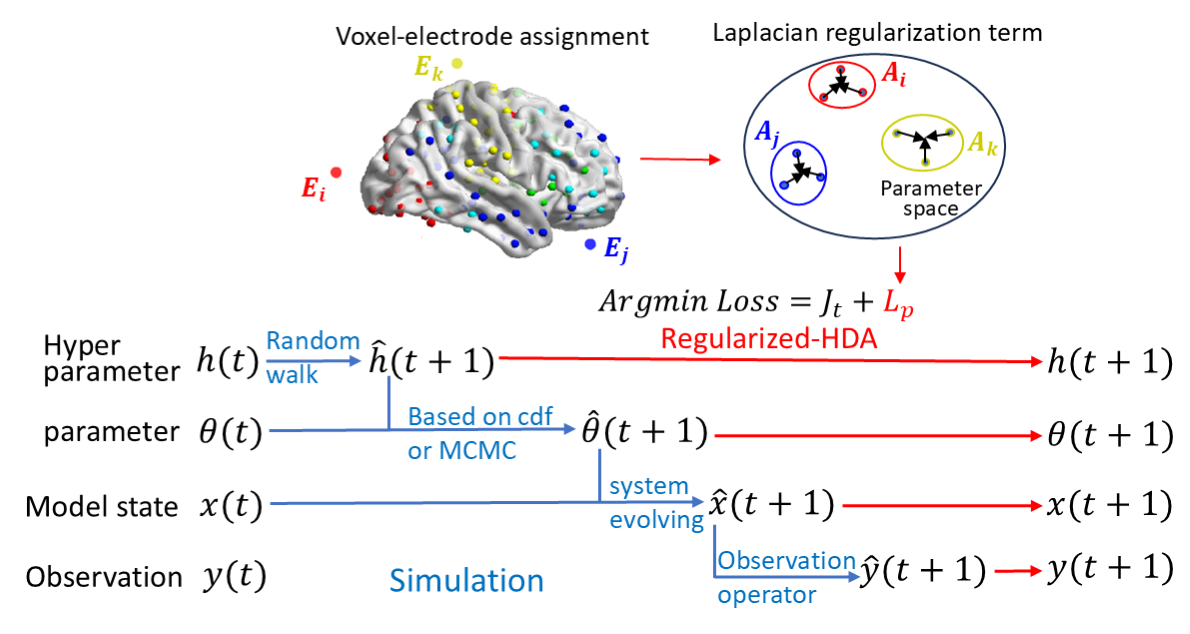}
  \caption{ {\bf Framework of Regularized-hierarchical data assimilation(Regularized-HDA).}The proposed Bayesian inference framework estimates hyperparameters by iterating two processes: simulation(blue arrows) and filtering(red arrows). During simulation, parameters are sampled from distributions determined by hyperparameters to propagate system states, with observations subsequently derived through observation operator; The filtering updates the hidden states by employing the Regularized-HDA algorithm, which introduces a Laplacian regularization term constructed from voxel-electrode assignments into the cost function.}
  \label{fig:fig2}
\end{figure}

Addressing the critical need for multimodal integration in computational neuroscience, we present a generalizable Bayesian framework integrating high-spatial-resolution MRI and high-temporal-resolution EEG data to construct biologically realistic brain models, where simulated EEG signals are decoded via EEGNet-LSTM to predict steering angles, ultimately enabling vehicle control in the CARLA simulator (Fig.~\ref{fig:fig}). Our contributions are: (1) We constructed an EEG-fused digital twin brain (EEG-DTB) that integrates structural MRI(sMRI)-derived cortico-subcortical model with empirical EEG data(Fig~\ref{fig:fig1} A); (2) We proposed a Bayesian inference framework based on the prior approaches~\cite{zhang2024framework, zhang2023deep} through the introduction of a Laplacian regularization term within the Kalman filter's cost function (Fig~\ref{fig:fig2}). This regularization, derived from the cortical voxel-to-electrode assignment, effectively resolved the ill-posed nature of the inverse problem, where over 10,000 hyperparameters were estimated from only 63 electrodes; (3) EEG-DTB's simulated signals were decoded via EEGNet-LSTM ~\cite{2023EEGNetLSTM, Lawhern2018}, producing steering predictions significantly superior to chance and empirical signals(p<0.05, Wilcoxon signed-rank test). Successful vehicle control in the CARLA simulator validated the model's ability to functionally replicate sensorimotor behavior (Fig~\ref{fig:fig1} C).

\section{Methods}
\label{headings}

\subsection{Data acquisition and preprocessing}

\subsubsection{Data acquisition}

All neuroimaging and electroencephalogram data were acquired from a single subject and provided by the Zhangjiang International Brain Imaging Center in Shanghai. Structural and diffusion-weighted images were scanned using a 3 Tesla MR scanner(Siemens Magnetom Prisma, Erlangen, Germany) equipped with a 64-channel head array coil. A high-resolution T1-weighted(T1w) image was acquired using a 3D magnetization-prepared rapid gradient echo(3D-MPRAGE) sequence. Additionally, multi-shell multi-band diffusion-weighted images(DWI) were obtained using a single-shot spin-echo planar imaging(EPI) sequence with a monopolar scheme. The scalp EEG data were captured using a 63-channel cap with a sampling rate of 1000 Hz and a low-pass filter of 280 Hz. During preparation, three cardinal points — the left preauricular point(LPA), the right preauricular point(RPA), and the nasion(NAS) — were recorded and subsequently used to co-register the EEG data with the MRI data. 

\subsubsection{Data preprocessing}

All EEG data were pre-processed using MNE-Python(v1.6.1, \href{https://doi.org/10.5281/zenodo.592483}
{https://doi.org/10.5281/zenodo.592483})~\cite{gramfort2013meg}. For resting-state data, bad segments were manually annotated and excluded, and noisy channels were interpolated based on surrounding
electrodes. The data were then re-referenced to the average reference. To repair low-frequency drift, a high-pass filter(0.01 Hz cutoff) was applied, followed by a low-pass filter(100 Hz cutoff) to
attenuate high-frequency noise. Power line interference was suppressed using notch filters centered at 50 Hz and its harmonics (0–100 Hz range). Finally, artifact removal was performed via the
independent component analysis(ICA) with 40 components, using the default ’fastica’ method ~\cite{hyvarinen1999fast}.

For task-state data, epochs were extracted from stimulus onset to offset. Similar to resting-state processing, bad segments were rejected, and bad channels were interpolated. After applying average 
re-referencing, the data were bandpass-filtered(0.5-80 Hz) and downsampled to 200 Hz. Notch filtering(50 Hz and harmonics) and ICA-based artifact removal(40 components, ’fastica’ method) were implemented.

The T1w and DWI data were processed following procedures similar to previous work~\cite{lu2024imitating}, with modifications to the brain mask generation based on forward modeling (see ~\nameref{MRI-constrained forward model}). The T1w data was preprocessed to obtain the voxel-based morphometry(VBM) of gray matter while structural connectivity matrices were derived from the DWI data (see ~\nameref{Supplementary materials}). After additional refinement for network modeling compatibility, 12,363 cortical and 1,951 subcortical voxels were retained in the network.

\subsubsection{Experimental protocol}

The participant operated a vehicle in the CARLA driving simulator using a physical steering wheel while wearing a 63-channel EEG cap. The participant used a physical steering wheel to control only the steering of a simulated vehicle. Speed was automatically regulated based on the vehicle’s angular velocity, allowing the participant to focus on maneuvering the vehicle.

Each trial consisted of two driving sessions: Outbound session – driving from a designated start to destination; Inbound session – returning to the original start point. To ensure focus on the designated route and minimize attentional bias, all alternative paths at junctions were physically blocked. Each session was preceded by a resting period, initiated via a steering wheel button press. Precisely 1-second fixation phase bracketed each driving session (before and after), serving both as participant cues and EEG epoch markers.

The participant completed four trials on an identical CARLA map using the same vehicle blueprint to ensure consistency in vehicle dynamics. Each trial consisted of paired outbound and inbound sessions(8 sessions total). The driving video stimulus was presented from three distinct viewpoints: (1) first-person (in-car) view, (2) third-person chase view, (3) top-down aerial view.

\subsection{Models}

\subsubsection{Whole-brain neuronal network model}

\paragraph{Computational neuronal model}

The computational neuron model is typically represented as a nonlinear operator that takes the presynaptic spike trains as input and generates the axon spike trains as output. Here, we adopted the leaky integrate-and-fire(LIF)~\cite{brunel2001effects} model(see ~\nameref{Supplementary materials}). In the EEG forward model, each cortical voxel was modeled as a current source, with the magnitude of the source current determined by the aggregate post-synaptic currents of all neurons within the voxel, i.e. :
\begin{eqnarray}
    \quad I_{v} &= \sum _{i\in v}I_{syn, i}, \quad v=1, 2, \cdots p
    \label{eq: source current(1)}
\end{eqnarray}
where $I_{v}$ represents the current magnitude of current source $v$, $p$ denotes the total number of current sources, $I_{syn,i}$ is the post-synaptic current of neuron $i$. Here, we considered four synapse types: AMPA, NMDA, $\mathrm{GABA_{A}}$, and $\mathrm{GABA_{B}}$ ~\cite{yang2016dendritic}.

\paragraph{Network topology}

Our whole-brain network model was constructed following the DTB framework~\cite{lu2022human}, comprising 12,363 cortical and 1,951 subcortical voxels (14,314 total). The whole brain was divided into 400 areas according to the HCPex parcellation~\cite{huang2022extended}, with structural connectivity probabilities between voxels derived from long-range fiber tracts identified through DWI-based tractography. Each cortical voxel was modeled as micro-columns containing eight populations~\cite{binzegger2004quantitative, PMID:23093925}, while each subcortical voxel was regarded as a canonical voxel with two populations. Across all voxels, we maintained a 4:1 ratio of excitatory to inhibitory neurons. Population-specific connection probabilities were implemented for intra-voxels. Finally, with a total of $N=10$ million neurons and a preset average in-degree $D=100$, we constructed a whole-brain neuronal network model including both cortical and subcortical structures. Complete implementation specifications are provided in ~\nameref{Supplementary materials}.

\subsubsection{MRI-constrained forward model}
\label{MRI-constrained forward model}

\paragraph{EEG forward model}

The EEG forward model calculates the scalp potential distribution generated by neural current sources ~\cite{Mosher1999}, requiring construction of a lead-field matrix. Mathematically, the forward model can be expressed as the inner product of a vector lead field and the source currents ~\cite{Tripp1983}. For $m$ electrodes and $p$ voxels, the forward model takes the following form:
\begin{eqnarray}
\mathbf{\Phi} = \mathbf{L} \cdot \mathbf{Q} = \begin{bmatrix}
\vec{\mathbf{L}}_{11}, \vec{\mathbf{L}}_{12}, \cdots,  \vec{\mathbf{L}}_{1p} \\
\vec{\mathbf{L}}_{21}, \vec{\mathbf{L}}_{22}, \cdots,  \vec{\mathbf{L}}_{2p}\\
\cdots, \cdots, \cdots, \cdots \\
\vec{\mathbf{L}}_{m1}, \vec{\mathbf{L}}_{m2}, \cdots, \vec{\mathbf{L}}_{mp}
\end{bmatrix} \cdot \begin{bmatrix}
\vec{\mathbf{Q_{1}} } \\
\vec{\mathbf{Q_{2}} }\\
\cdots \\
\vec{\mathbf{Q_{p}} }
\end{bmatrix}
\label{eq: MEG forward model (1)}
\end{eqnarray}
Where, $\mathbf{\Phi} \in \mathbb{R}^{m \times 1}$ represents the EEG signals recorded by $m$ electrodes, $\mathbf{L} \in \mathbb{R}^{m \times 3p}$ is the lead-field matrix, $\mathbf{Q} \in \mathbb{R}^{3p \times 1}$ denotes the source currents of all voxels. $\vec{\mathbf{L}_{iv}} \in \mathbb{R}^{1 \times 3}$ represents the vector lead field relative to electrode $i$ and voxel $v$. $\vec{\mathbf{Q}_{v}} \in \mathbb{R}^{3 \times 1}$ represents the source current of voxel $v$, defined as :
\begin{equation}
\vec{\mathbf{Q}_{v}} = I_{v} \cdot \vec{\mathbf{p}}_{v}
\label{eq: source current}
\end{equation}
where $\vec{\mathbf{p}}_v \in \mathbb{R}^{3 \times 1}$ is the unit vector determined by positions of voxel $v$ and those voxels form connections with voxel $v$.

The lead-field matrix was calculated following a standardized processing pipeline implemented in Brainstorm (\href{http://neuroimage.usc.edu/brainstorm}{http://neuroimage.usc.edu/brainstorm}), which integrated the individual's T1w MRI and EEG data. We employed the Boundary Element Method(BEM) from OpenMEEG software ~\cite{Gramfort2010} to solve the forward model, with focus on cortical voxels. The voxel positions derived from Brainstorm's anatomical processing served two critical purposes: (1) Generation of a brain mask for network construction; (2) Determination of source current orientation vectors ($\vec{\mathbf{p}}_{v}$) based on both voxel positions and inter-voxel connectivity patterns.

\paragraph{MRI-constrained forward model}

Let $M \in \mathbb{R}^{m \times p}$ be the sensitivity matrix defined by:
\begin{eqnarray*}
    \begin{array}{c}
M = \begin{bmatrix}
M_{11}, M_{12}, \cdots , M_{1p}\\
M_{21}, M_{22}, \cdots , M_{2p}\\
\cdots, \cdots , \cdots , \cdots \\
M_{m1}, M_{m2}, \cdots , M_{mp}
\end{bmatrix}
\end{array}, 
 \quad M_{i,v} = \vec{\mathbf{L}}_{i,v} \cdot \vec{p}_{v}
\end{eqnarray*}
Then the forward model Eq(~\ref{eq: MEG forward model (1)}) becomes:
\begin{eqnarray}
     \mathbf{\Phi} = \begin{bmatrix}
\sum_{v=1}^{p}  M_{1,v}I_{v},\\
\sum_{v=1}^{p}  M_{2,v}I_{v}, \\
\cdots \\
\sum_{v=1}^{p}  M_{m,v}I_{v}
\end{bmatrix} 
\label{eq: MEG forward model (2)}
\end{eqnarray}
From Eq(~\ref{eq: MEG forward model (2)}), $|M_{i,v}|$ represents the sensitivity of electrode $i$ to activity of voxel $v$. Thus, we optimally assigned cortical voxels to electrodes by solving:
\begin{equation}
\underset{\mathcal{A}}{\text{maximize}} \sum_{i=1}^m \sum_{v \in \mathcal{A}_{i}} |\mathbf{M}_{i,v}|
\end{equation}
using the Hungarian Algorithm~\cite{kuhn1955hungarian}, where $\mathcal{A}i$ denotes voxels allocated to electrode $i$. The MRI-constrained forward model was obtained by modifying the lead-field matrix as follows:
\begin{equation}
\tilde{\vec{\mathbf{L}}}_{i,v} = \left\{\begin{matrix}
 \mu \vec{\mathbf{L}}_{i,v}, \quad if \quad v \notin \mathcal{A}_{i}\\ 
\vec{\mathbf{L}}_{i,v}, \quad if \quad v \in \mathcal{A}_{i}
\end{matrix}\right.
\end{equation}
where $\mu \in (0, 1]$ is a shrinkage coefficient for non-allocated voxels.

\subsection{Bayesian inference framework}

We proposed a Regularized-hierarchical data assimilation(Regularized-HDA) algorithm to estimate the parameters of our neuronal network model based on empirical EEG signals. Building upon the HDA framework ~\cite{zhang2024framework} incorporating the diffusion strategy ~\cite{zhang2023deep}, we introduced a Laplacian regularization term derived from the voxel-electrode assignment $\mathcal{A}$ obtained in the previous section(see Fig~\ref{fig:fig2}). The HDA framework assumes that the parameters of a pool of neurons are independently and identically sampled from the distribution determined by the same hyperparameter. Rather than directly estimating individual neuron parameters, we inferred these hyperparameters. The diffusion strategy enhances this process by weighting the estimates from electrode observations through the fusion coefficient $\gamma$, where hyperparameters for voxels in $\mathcal{A}_{i}$ are preferentially influenced by measurements from electrode $i$. To address the ill-posedness of the inverse problem, where the dimension of hyperparameters surpasses the number of electrodes, we introduced a geometrically-constrained regularization term expressed as:
\begin{equation}
L_{p}=\sum_{i=1}^{m} \sum_{j \in \mathcal{A}_{i} }|h_{j} - \bar{h}_{i}|^{2}  \label{eq:Laplace regularization}
\end{equation}
where:
\begin{equation}
    \bar{h}_{i} = \frac{\sum_{j\in \mathcal{A}_{i}} h_{j}}{\# \mathcal{A}_{i}}
\end{equation}
denotes the mean of the hyperparameters of voxels $\mathcal{A}_{i}$. The regularization term $L_{p}$ was incorporated into the Kalman filter's cost function. By setting the derivative of the cost function with respect to the Kalman gain matrix to zero, we obtained the updated formula for the Kalman gain matrix:
\begin{eqnarray}
    K = (I + \lambda  \mathcal{L})^{-1}\left[\hat{P} H^ \top  - \lambda \mathcal{L} \mathbb{E}\left[\hat{x}(y-H\hat{x})^ \top \right]\right](H\hat{P} H^\top+\Gamma)^{-1}
    \label{eq: new Kalman gain matrix}
\end{eqnarray}
where $\lambda > 0$ is the penalty coefficient, $\mathcal{L} \succeq  0$ is the penalty matrix derived from the regularization term, $H$ is the matrix mapping the state variables to the observations, $\hat{x}$ represents the prior estimate of the state variables, $\hat{P}$ is the prior estimate error covariance matrix of state variables, $y$ denotes the observations at a given time point, $\Gamma$ represents the observation noise covariance matrix, $I$ is the identity matrix. We replaced the Kalman gain matrix computation in ~\cite{zhang2024framework, zhang2023deep} with the regularized formulation Eq(~\ref{eq: new Kalman gain matrix}). The complete algorithm was detailed in ~\nameref{Supplementary materials}.

\subsection{EEGNet-LSTM}
\label{EEGNet-LSTM}

To decode driving behavior events - specifically, the steering angle - from EEG signals, we employed a hybrid neuronal network model, EEGNet-LSTM~\cite{2023EEGNetLSTM, Lawhern2018}. The model architecture comprises two key components(see Fig~\ref{fig:fig1} C): (1)EEGNet, a convolutional neural network(CNN) for front-end spatial-frequency feature extraction, and (2) an LSTM network to capture temporal dependencies in the extracted features. Building upon the regression model architecture in Table 2 of ~\cite{Lawhern2018}, we presented our network structure in ~\nameref{Supplementary materials}.

The temporal synchronization was achieved by precisely aligning the EEG data with the timestamps of CARLA simulator outputs, such as the steering angle. We segmented the continuous EEG signals into 1-second patches (200 time points at a 200 Hz sampling rate) synchronized with behavioral event markers, then paired them with corresponding steering angles for EEGNet-LSTM training.

We trained the model in PyTorch using cross-validation (3 outbound sessions for training, 1 for testing), Adam optimization (lr=0.001), and mean square error(MSE) loss for steering angle prediction.

\subsection{Evaluation metric}

We used the Pearson correlation coefficient(PCC) and mean relative square error(MRSE) between simulated and empirical EEG signals to evaluate Bayesian inference performance, and MSE and mean absolute error(MAE) between predicted and true steering angles to assess EEG decoding.

\section{Results}

\subsection{EEG-DTB in resting-state}

For the resting-state EEG-DTB model(see Fig~\ref{fig:fig1} A), we implemented two key configurations: (1) the NMDA synaptic conductances of neurons were modified, and the conductances of neurons in each voxel follow the Gamma distribution with a shape parameter $\alpha = 5$ and an inverse scale parameter $\beta = 5 / g_{h}$(where $g_{h}$ is the voxel-specific hyperparameter); (2) all external currents were set to zero. The hyperparameters of all voxels were estimated through fitting empirical resting-state EEG signals under millisecond resolution.

\begin{figure}
  \centering
  \includegraphics[width=5in]{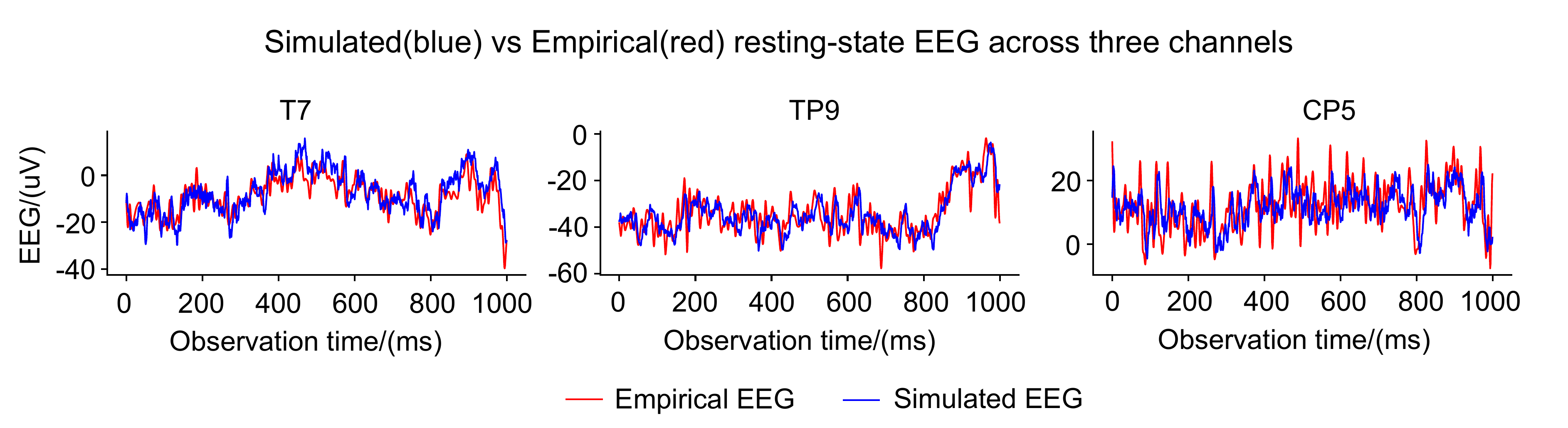}
  \caption{ {\bf Resting-state EEG-DTB simulation.} Comparison of simulated (blue) and empirical (red) resting-state EEG signals across three channels, demonstrating strong agreement (PCC=0.88).}
  \label{fig:fig4}
\end{figure}

The Regularized-HDA algorithm was implemented with specific parameter configurations: the shrinkage coefficient in the MRI-constrained forward model was set to $\mu = 0.1$, while the fusion and penalty coefficients in the Regularized-HDA algorithm were assigned $\gamma = 0.5$ and $\lambda = 0.001$, respectively. Additional algorithmic parameters are detailed in ~\nameref{Supplementary materials}. The total duration of the simulation is $T = 10 s$ with an iterative time step of 1 ms. Fig ~\ref{fig:fig4} compares the simulated(blue) and empirical(red) EEG signals across three electrodes, demonstrating strong agreement with a PCC of 0.88 and MRSE of 0.007.

\subsection{EEG-DTB in action}

Building upon the resting-state EEG-DTB, we administered stimuli to visual processing areas comprising the primary visual cortex (V1), early visual cortices (V2/V3), dorsal and ventral streams, and prefrontal regions. The stimulus patterns were derived from task-state EEG data assimilation, generating time-varying external currents that were applied to excitatory populations. Specifically, these currents followed a Gamma distribution with shape parameter $\alpha = 5$ and inverse scale parameter $\beta = 5/I_{h}$, where $I_{h}$ represents the voxel-specific hyperparameter.

\begin{figure}
  \centering
  \includegraphics[width=5.5in]{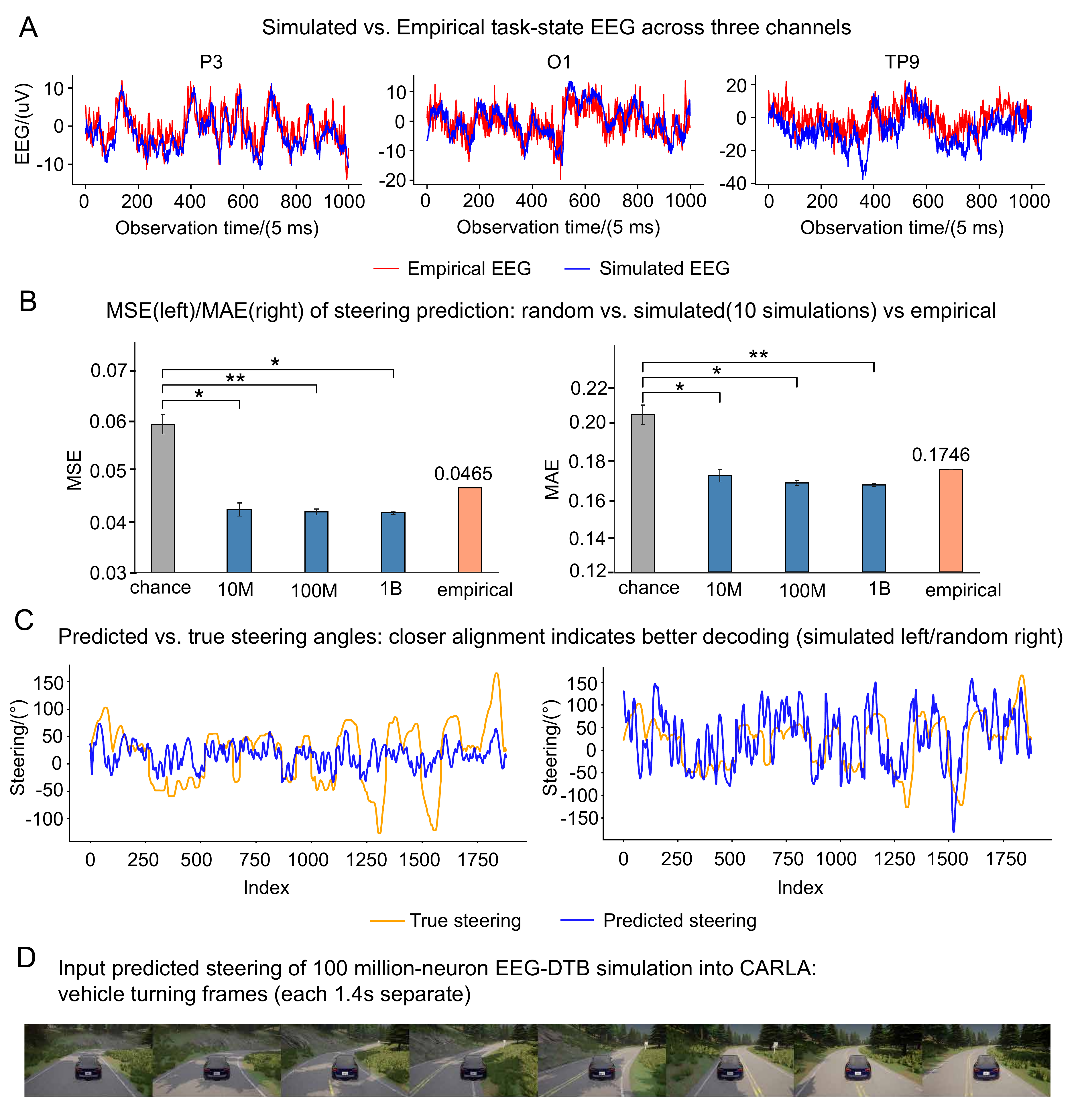}
  \caption{ {\bf Task-state EEG-DTB: vehicle driving control in virtual scenarios.} (A) Comparison of simulated (blue) and empirical (red) task-state EEG signals across three channels, with a PCC of 0.60. (B) MSE(left) and MAE(right) of steering prediction. We performed 10 simulation trials per network scale and 10 random permutations of empirical signals. The MSE and MAE between predicted and true steering angles were calculated after normalizing both values by $450^{\circ}$ (representing 1.5 full steering wheel rotations). Decoding steering revealed significantly lower MSE(left) and MAE(right) values for simulated signals compared to both random(p < 0.05, Wilcoxon test) and empirical signals(95 \% confidence interval for MSE/MAE across models: 10M [0.0414,0.0432]/[0.1688,0.1736], 100M [0.0414,0.0422]/[0.1665,0.1684], 1B [0.0414,0.0419]/[0.1658,0.1669]). (C) Steering angle prediction for EEG-DTB-generated signals (left) versus random signals (right), with predicted (blue) and true (orange) angles aligned by timestamps. The horizontal axis labels indicate the sequential index of steering angles sorted by their corresponding timestamps. (D) Representative frames(each 1.4 s separate) of the vehicle driving through curves in the CARLA simulator.}
  \label{fig:fig5}
\end{figure}

The Regularized-HDA algorithm maintained identical parameter configurations to those used in resting-state assimilation. We performed task-state EEG signal assimilation across 31,155 time points, spanning the complete session duration from onset to offset. Throughout the simulation, post-synaptic currents were computed and projected to EEG via our MRI-constrained forward model(Fig~\ref{fig:fig1} B). To align with empirical recordings, the simulated signals were downsampled to match the 200 Hz downsampling rate of the experimental EEG data. Fig ~\ref{fig:fig5} A shows the simulated(blue) and empirical(red) EEG signals of three electrodes, with a PCC of 0.52 and MRSE of 1.89.

We then systematically increased the number of neurons in the network while maintaining identical architecture and hyperparameters. Quantitative comparisons between simulated and empirical signals across different network scales (Table ~\ref{tab:network scale}) reveal that the agreement between simulated and empirical signals improves with network size.

\begin{table}
  \caption{PCC and MRSE for different numbers of neurons in the network model}
  \label{tab:network scale}
  \centering
  \begin{tabular}{llll}
    \toprule                   
    Number of neurons     & 10 million     & 100 million & 1 billion \\
    \midrule
    PCC & 0.52  & 0.59 & 0.60   \\
    MRSE & 1.89 & 1.77 & 1.84    \\
    \bottomrule
  \end{tabular}
\end{table}

\subsection{EEG decoding via EEGNet-LSTM model}

We trained an EEGNet-LSTM model using four-fold cross-validation to predict steering angles from EEG signals. Both simulated and empirical EEG signals were segmented as 1-second patches synchronized with steering angles. The decoding analysis encompassed three comparisons: simulated EEG signals generated by EEG-DTB models of different network scales, original empirical EEG recordings, and random permuted empirical signals. We performed 10 simulation trials per network scale and 10 random permutations of empirical signals. The MSE and MAE between predicted and true steering angles were calculated after normalizing both values by $450^{\circ}$ (representing 1.5 full steering wheel rotations) (Fig.~\ref{fig:fig5}B). Fig ~\ref{fig:fig5} C shows the predicted(blue) and true(orange) steering angles for simulated(left) and random(right) signals, respectively. 

Fig.~\ref{fig:fig5} B quantitatively demonstrates the superior steering angle prediction of our EEG-DTB's simulated signals, showing significantly lower MSE/MAE than both: (1) random signals (p<0.05, Wilcoxon test) and (2) empirical EEG recordings (95 \% CIs for 10M/100M/1B models: MSE=[0.0414-0.0432]/[0.0414-0.0422]/[0.0414-0.0419]; MAE=[0.1688-0.1736]/[0.1665-0.1684]/[0.1658-0.1669]). Notably, the EEG-DTB model with 100 million and 1 billion neurons generated simulated signals that demonstrated improved decoding accuracy compared to the 10 million-neuron model, suggesting that larger-scale network simulations better capture the neural dynamics underlying complex sensorimotor mechanisms.

Finally, we input the predicted steering angles of the simulated signals generated by the EEG-DTB model with 100 million neurons into the CARLA simulator. Given the starting point and end point, the vehicle successfully navigated through curves using the predicted steering angles, with an adaptive speed regulation that reduced velocity for larger steering wheel deflection angles. Fig ~\ref{fig:fig5} D shows representative frames of this autonomous curve negotiation, while a complete driving demo is available in the video \href{https://anonymous.4open.science/r/EEG-fused-Digital-Twin-Brain-for-Autonomous-Driving-in-Virtual-Scenarios-B43B/}{https://anonymous.4open.science/r/EEG-fused-Digital-Twin-Brain-for-Autonomous-Driving-in-Virtual-Scenarios-B43B/}.

Our results demonstrate that the EEG-DTB model successfully emulates the complete sensorimotor processing chain observed in biological systems - from perceptual input reconstruction to neural encoding and ultimately behavioral output. This replication provides evidence that the model captures essential information flow characteristics along the canonical visual-motor pathway.

\section{Conclusion}

Our study developed a Bayesian inference framework integrating MRI with EEG data to construct a biologically realistic DTB model. The EEG-DTB demonstrated strong agreement with empirical recordings, achieving a correlation of 0.88 for resting-state (Fig ~\ref{fig:fig4}) and 0.60 for task-state EEG (Fig ~\ref{fig:fig5} A, Table ~\ref{tab:network scale}). Furthermore, the model's simulated task-state signals achieved significantly better steering decoding than both random and empirical signals (p<0.05, Wilcoxon signed-rank test), enabling successful vehicle control in the CARLA simulator(Fig~\ref{fig:fig5} D). This demonstrates a complete brain simulation-to-behavior pipeline.

We established voxel-electrode assignments through the EEG forward model derived from T1w anatomy and electrode geometry to bridge the spatial resolution disparity between MRI and EEG data. This anatomical constraint was formalized as a Laplacian regularization term added to the Kalman filter's cost function, effectively solving the ill-posed inverse problem of estimating over 10,000 hyperparameters from only 63 EEG channels. Using our Regularized-HDA framework to assimilate EEG signals into the DTB model constructed from sMRI data, we implemented vehicle control in the CARLA simulator via EEG signals generated by the EEG-DTB model. This approach fundamentally differs from conventional end-to-end deep learning systems by establishing a direct, interpretable linkage between decoded driving behavior and the underlying spiking neuronal dynamics of the EEG-DTB model. 

Our EEG-DTB model functionally replicated three core sensorimotor processes: (1) sensory input reconstruction through EEG-derived currents, (2) neural encoding via EEG generation and biologically constrained network simulations, and (3) behavior generation by decoding steering. This pipeline—from sensory stimulus reconstruction to behavioral control—demonstrates that our model captures fundamental encoding and decoding principles underlying biological sensorimotor processing. 

Our EEG-DTB model provides a unique platform to decode information flow dynamics: encoding, transmission, and integration, along neural pathways. By selectively stimulating specific brain regions with EEG-derived currents and analyzing network-wide activation patterns, researchers could characterize how hierarchical transformations occur along the visual-dorsal-motor pathway.

Building upon the current framework, future work will enhance the EEG-DTB's behavioral capabilities by implementing modulation of long-range connectivity weights. This refinement will leverage task-state neural signatures to optimize information routing through cortico-cortical and cortico-subcortical pathways, potentially improving the model's sensorimotor transformation accuracy for complex tasks.


\bibliographystyle{unsrt}

\newpage

\appendix

\setcounter{figure}{0}
\renewcommand{\thefigure}{A.\arabic{figure}}
\setcounter{table}{0}
\renewcommand{\thetable}{A.\arabic{table}}
\setcounter{equation}{0}
\renewcommand{\theequation}{A.\arabic{equation}}
\renewcommand{\theenumiv}{A\arabic{enumiv}} 
\setcounter{enumiv}{0} 

\section{Technical Appendices and Supplementary Material}
\label{Supplementary materials}

\subsection{Computational neuronal model}

In this work, we adopted the leaky integrate-and-fire(LIF)~\cite{Abrunel2001effects} model, governed by the following evolution law:
\begin{eqnarray}
\left\{\begin{array}{l}
C_{i} \dot{V}_{i}=-g_{L}\left(V_{i}-V_{L}\right)+I_{syn, i}+I_{bg, i}+I_{ext, i}, \quad V_{i}<V_{t h} \\
I_{syn, i} = \sum_{u} I_{u, i} = \sum_{u}g_{u, i}\left(V_{u}-V_{i}\right) J_{u, i}, \\
\dot{J}_{u, i}=-\frac{J_{u, i}}{\tau_{i}^{u}}+\sum_{k, j} w_{i j}^{u} \delta\left(t-t_{k}^{j}\right), \\
V_{i}(t)=V_{\text {rest }}, \quad t \in\left(t_{k}^{i}, t_{k}^{i}+T_{ref}\right] .
\end{array}\right.
\label{eq: CV equation}
\end{eqnarray}
where $C_{i}$ represents the capacitance of the neuron membrane, $g_{L}$ denotes the leakage conductance, $V_{L}$ is the leakage voltage, and $I_{syn, i}$ refers to the synaptic current of neuron $i$, defined as the sum of currents of four synapse types. $I_{u, i}$ represents the synaptic current for synapse type $u$ of neuron $i$. Here, we considered four synapse types: AMPA, NMDA, $\mathrm{GABA_{A}}$, and $\mathrm{GABA_{B}}$ ~\cite{Ayang2016dendritic}. $g_{u, i}$ denotes the conductance of synapse type $u$ for neuron $i$, which can be modified during data assimilation. $V_{u}$ is the voltage of synapse type $u$, $\tau_{i}^{u}$ is the time-scale constant of synapse type $u$ for neuron $i$. $w_{ij}^{u}$, representing the connection weights from neuron $j$ to $i$ of synapse type $u$, are drawn from the uniform distribution $U(0, 1)$. $\delta\left(\cdot\right)$ is the Dirac-delta function and $t_{k}^{j}$ is the time point of the $k-th$ spike of neuron $j$. $I_{ext, i}$ denotes the external stimulus applied exclusively to excitatory neurons. In the resting state, $I_{ext, i}=0$. The background current, $I_{bg,i}$, was modeled as independent Ornstein-Uhlenbeck(OU) processes, defined as follows:
\begin{eqnarray}
    \tau_{b g} d I_{b g, i}=\left(\mu_{b g}-I_{b g, i}\right) \mathrm{dt}+\sqrt{2 \tau_{b g}} \sigma_{b g} d W_{t} .
    \label{eq: OU current}
\end{eqnarray}
where $\mu_{bg}$ and $\sigma_{bg}$ represent the mean and standard deviation of the background current, respectively. $\tau_{bg}$ denotes the time-scale constant and $W_{t}$ is the Wiener process. When the membrane potential of neuron $i$, $V_{i}$, reaches a given voltage threshold $V_{th}$, the neuron emits a spike, and $V_{i}$ is reset at $V_{rest}$ during a refractory period $T_{ref}$. After that, $V_{i}$ evolves according to the first equation of Eq~(\ref{eq: CV equation}) again. All constant parameters in the computational neuron model were set in Table~\ref{tab: si default parameters}.

\begin{table}[!ht]
  \centering
  \caption{{\bf Default parameters in the computational neuron model}}
  \label{tab: si default parameters}
  \begin{tabular}{ccccccccccc}
    \toprule
    {\bf Symbol} & {\bf Value}   \\
    \toprule
    $C_{i}$ & $1 \mu f $  \\
    \midrule
    $g_{L,i}$ & $0.03 mS$ \\
    \midrule
    $V_{L}$ & $-75 mV$ \\
    \midrule
    $V_{th}$ & $-50 mV$ \\
    \midrule
    $V_{rest}$ & $-65 mV$ \\
    \midrule
    $\left [ V_{AMPA, i}, V_{NMDA,i}, V_{GABAa,i}, V_{GABAb,i} \right ]$ & $\left [ 0, 0, -70, -100 \right ] mV$ \\
    \midrule
    $\left [ \tau_{AMPA, i}, \tau_{NMDA,i}, \tau_{GABAa,i}, \tau_{GABAb,i} \right ]$ & $\left [ 2, 40, 10, 50 \right ]$ \\
    \midrule
    $w_{i,j}^{u}$ &  $\sim U(0,1)$ \\
    \midrule
    $T_{ref}$ & $5 ms$ \\
    \midrule
    $\mu_{bg}$ & $0.6 nA$ \\
    \midrule
    $\sigma_{bg}$ & $0.2 nA$ \\
    \midrule
    $\tau_{bg}$ & $10 ms$ \\
    \bottomrule
  \end{tabular}
\end{table}

\subsection{Network topology}

We provided a brief overview of the digital twin brain(DTB) ~\cite{Alu2022human} architecture. For the cortex, each voxel was modeled as a micro-column mimicking cat primary visual cortex neuroanatomy~\cite{Abinzegger2004quantitative, APMID:23093925}, with layers L2/3-L6 divided into two populations each. The L1 layer was not considered due to the relatively small number of neurons it contains. Subcortical voxels adopted a canonical voxel structure with two populations. Given a predefined scale, the number of neurons per population was determined by the voxel-based morphometry(VBM) of the gray matter and layer-specific distributions provided by Du et al.~\cite{APMID:23093925}. The ratio between excitatory and inhibitory neurons was layer-specific and roughly 4:1 on average.

In this work, we assumed a constant average in-degree of 100 for neurons in each voxel. The structural connectivity probability matrix, derived from long-range fiber tracts identified through DWI-based tractography, determined inter-voxel connection probabilities. We classified synaptic connections into two types: (1) local connections within voxels and (2) long-range connections between voxels. Additionally, a key hypothesis in our model was that inter-voxel long-range connections were purely excitatory. For intra-voxel (local) connections between populations, we employed population-specific connectivity statistics. For cortical voxels, the connections between populations were derived from cat primary visual cortex anatomy ~\cite{Abinzegger2004quantitative,Afalchier2002anatomical} (Table~\ref{tab: connectivity matrix in micro-column}). The connectivity matrix presented in Table~\ref{tab: connectivity matrix in micro-column} represents the average number of synapses received by each post-synaptic neuron from the corresponding population. Specifically, each row represents a post-synaptic neuron type, where e1(i1) indicates the excitatory (inhibitory) neuron of layer 1, and so on. Each column represents a pre-synaptic neuron type, where CC indicates that the pre-synaptic neuron is in other voxels, respectively. For subcortical voxels, the connections between populations were given by Table~\ref{tab: connectivity matrix in subcortex}. For each target population, we calculated inter- and intra-voxel synaptic connection counts using Table~\ref{tab: connectivity matrix in micro-column} (cortex) and Table~\ref{tab: connectivity matrix in subcortex} (subcortex). Inter-voxel connections were allocated to excitatory source populations according to: (1) structural connectivity probabilities and (2) source population neuron counts. Combining these with the predefined scale and average in-degree (D=100), we constructed the whole-brain network model.

\begin{table}[!ht]
  \centering
  \caption{\bf {Connectivity matrix within micro-columns.}}
  \label{tab: connectivity matrix in micro-column}
  \begin{tabular}{c|c|cccccccccc}
    \toprule
    \multicolumn{12}{c}{Cortex} \\
    \midrule
  &  & \multicolumn{10}{c}{pre-synaptic}   \\
    \midrule
  &  & i1 & e2/3 & i2/3 & e4 & i4 & e5 & i5 & e6 & i6 & CC  \\
    \midrule  
 \multirow{10}{*}{\rotatebox{90}{post-synaptic}} & e1 & 1323 & 823 & 200 & 15 & 1 & 9 & & & & 10133 \\
   & i1 & 901 & 560 & 149 & 10 & 1 & 6 & & & & 6899 \\
   & e2/3 & 133 & 3554 & 804 & 881 & 45 & 431 & & 136 & & 1020 \\
   & i2/3 & 52 & 1778 & 532 & 456 & 29 & 217 & & 69 & & 396 \\
   & e4 & 27 & 417 & 84 & 1070 & 782 & 79 & 8 & 1686 & & 1489 \\
   & i4 &  & 168 & 41 & 628 & 538 & 36 & & 1028 & & 790 \\
   & e5 & 147 & 2550 & 176 & 765 & 99 & 621 & 596 & 363 & 7 & 1591 \\
   & i5 &  & 1357 & 76 & 380 & 32 & 375 & 403 & 129 & & 214 \\
   & e6 & 2 & 643 & 46 & 549 & 196 & 327 & 126 & 925 & 597 & 2609 \\
   & i6 &  & 80 & 8 & 92 & 3 & 159 & 11 & 76 & 499 & 1794 \\
    \bottomrule
  \end{tabular}
  \begin{flushleft} Average number of synapses received by individual neuron in each cortical layer. e1, e2/3, e4, e5, e6, and i1, i2/3, i4, i5, i6 represent the excitatory and inhibitory population of layer 1, 2/3, 4, 5, 6 respectively. 'CC' indicates that the pre-synaptic neuron originates from other voxels.
\end{flushleft}
\end{table}

\begin{table}[!ht]
  \centering
  \caption{{\bf Connectivity matrix within subcortical voxels.}}
  \label{tab: connectivity matrix in subcortex}
  \begin{tabular}{c|c|ccc}
    \toprule
    \multicolumn{5}{c}{Subcortex} \\
    \midrule
  &  & \multicolumn{3}{c}{pre-synaptic}   \\
    \midrule
  \multirow{6}{*}{\rotatebox{90}{post-synaptic}} &  & e & i & CC  \\  
  & & & & \\
 & e & $\frac{4}{7}$ & $\frac{1}{7}$ & $\frac{2}{7}$ \\
  & & & & \\
   & i & $\frac{4}{7}$ & $\frac{1}{7}$ & $\frac{2}{7}$ \\
   & & & & \\
    \bottomrule
  \end{tabular}
  \begin{flushleft} The ratio of synapses received by individual neuron in the subcortex. e and i represent the excitatory and inhibitory population, respectively. 'CC' indicates that the pre-synaptic neuron originates from other voxels.
  \end{flushleft}
\end{table}

\subsection{Data acquisition and preprocessing}

\subsubsection{Data acquisition}

In this work, all neuroimaging and electroencephalogram data were acquired from a single subject. Structural and diffusion-weighted images were scanned using a 3 Tesla MR scanner(Siemens Magnetom Prisma, Erlangen, Germany) equipped with a 64-channel head array coil. A high-resolution T1-weighted(T1w) image was acquired using a 3D magnetization-prepared rapid gradient echo(3D-MPRAGE) sequence with the following parameters: repetition time(TR)=3000ms, echo time(TE)=2.5 ms, inversion time (TI)=1100ms, field of view(FOV)=256mm, flip angle=$7^{\circ}$, matrix size= $320 \times 320$, 240 sagittal slices, slice thickness=0.8mm, and no gap. Additionally, multi-shell multi-band diffusion-weighted images(DWI) were obtained using a single-shot spin-echo planar imaging(EPI) sequence with a monopolar scheme. The imaging parameters were as follows: TR=3200ms, TE=82ms, matrix size=$140 \times 140$, voxel size=$1.5 \times 1.5 \times 1.5$ mm3, multiband factor=4, phase encoding direction = anterior-to-posterior, and two b-values of 1500 s/(30 diffusion directions) and 3000 s/(60 diffusion directions). 

\subsubsection{Data preprocessing}

A brain mask, including both cortex and subcortex, was generated using the voxel coordinates in the individual MRI space derived from EEG forward modeling. This mask was subsequently used in the following T1w and DWI image preprocessing.

As regards the voxel-based morphometry(VBM) of the T1w images, we employed the VBM8 toolbox in the Statistical Parametric Mapping package(SPM, \href{http://www.fil.ion.ucl.ac.uk/spm}{http://www.fil.ion.ucl.ac.uk/spm}). Specifically, the gray matter image was segmented in the individual MRI space, smoothed with a full-width at half-maximum(FWHM) 8-mm Gaussian kernel, and then resampled to a resolution of $0.8 \times 0.8 \times 0.8$ mm3.

We utilized FSL software V6.0.4(Functional Magnetic Resonance Imaging of the Brain Software Library, \href{http://www.fmrib.ox.uk/fsl}{http://www.fmrib.ox.uk/fsl}) and MRtrix 3.0 (\href{http://www.mrtrix.org}{http://www.mrtrix.org})~\cite{Atournier2019mrtrix3} to preprocess the DWI data. First, the DWI data was denoised and corrected for Gibbs ringing artifact~\cite{Averaart2016denoising, Averaart2016diffusion, Acordero2019complex, Akellner2016gibbs}, followed by correction for head motion and eddy currents using the reversed phase encoding $b=0 s/ mm2$ images~\cite{Aandersson2003correct, Aandersson2016incorporating, Aandersson2016integrated, Aleemans2009b}. Subsequently, the bias field correction was applied to the DWI data~\cite{Azhang2001segmentation, Asmith2004advances}. To obtain the anatomical connectivity matrix, we employed the DWI-based tractography. Briefly, we first estimated the number of fiber tracts connecting each pair of voxels in the brain mask mentioned above through white matter tractography. The structural connectivity matrix was then derived from the row-normalized fiber tract counts. In the connectivity matrix, the connection of a voxel to itself was set to 0. 

To better align the neuroimaging data with our neuronal network model, we performed further cleaning on the preprocessed data based on the following principles: (1) For cortical structures, voxels with a gray matter volume(GMV) less than 0.3 were discarded, while for subcortical structures, voxels with a GMV less than 0.2 were excluded; (2) Isolated voxels that were not connected with any other voxels were removed. Ultimately, 12,363 cortical and 1,951 subcortical voxels were retained in the network.

\subsection{Bayesian inference framework}

\subsubsection{A general form of Kalman Filter with quadratic regularization term}

First, we gave the derivation of the general form of the Kalman filter incorporating a quadratic regularization term.

Consider the following linear dynamical system with Gaussian white noise:
\begin{eqnarray}
    \begin{array}{l}
x_{t}=F_{t-1} x_{t-1}+G_{t-1} u_{t-1}+\xi _{t-1} \\
y_{t}=H_{t} x_{t}+\eta _{t}
\end{array}
\label{eq: linear dynamic system}
\end{eqnarray}
where $x_{t}$ represents the state variables of the system, and $y_{t}$ denotes the system's observations. The term $\xi_{t-1}$ and $\eta_{t}$ are Gaussian white noise with covariance matrices $Q_{t-1}$ and $R_{t}$, respectively, where $\xi_{t-1}$ is referred to as the model noise and $\eta_{t}$ the observation noise. The matrix $F_{t-1}$ is the state transition matrix describing the system's evolution from time step $t-1$ to $t$, while $u_{t-1}$ represents the control input vector, and $G_{t-1}$ is the control input matrix that relates the control input vector to the state. The matrix $H_{t}$ maps the state $x_{t}$ to the observations $y_{t}$ at time step $t$. Given the true observations $y$ at each step, the Kalman filter computes the optimal estimate of the state variables in the sense of minimizing the mean-square error(MSE).

The Kalman filter consists of two steps: the prediction step and the analysis step. During the prediction step, the system evolves according to Eq~(\ref{eq: prediction step}) to obtain the prior estimate and the prior estimate error covariance:
\begin{eqnarray}
    \begin{array}{l}
    \hat{x}_{t}^{-}=F_{t-1} \hat{x}_{t-1}^{+}+G_{t-1} u_{t-1} \\
    P_{t}^{-}=F_{t-1} P_{t-1}^{+}F_{t-1}^\top+ Q_{t-1}
    \end{array} 
    \label{eq: prediction step}
\end{eqnarray}
here, the superscripts $-$ and $+$ denote the prediction(prior) and update(posterior) of the corresponding variables, respectively. 

For the analysis step, given the true observations $y_{t}$ and the prior estimate $\hat{x}_{t}^{-}$, the posterior estimate $\hat{x}_{t}^{+}$ and the posterior estimate error covariance $P_{t}^{+}$ is given by:
\begin{eqnarray}
    \begin{aligned}
        K_{t}&=P_{t}^{-}H_{t}^\top(R_{t}+H_{t}P_{t}^{-}H_{t}^\top)^{-1} \\
        \hat{x}_{t}^{+}&= \hat{x}_{t}^{-} + K_{t}(y_{t}-H_{t}\hat{x}_{t}^{-}) \\
        P_{t}^{+}&=\left(I-K_{t} H_{t}\right) P_{t}^{-}\left(I-K_{t} H_{t}\right)^\top+K_{t} R_{t} K_{t}^\top=(I-K_{t}H_{t})P_{t}^{-}
    \end{aligned}
    \label{eq: analysis step}
\end{eqnarray}
where $K_{t}$, known as the Kalman gain matrix, is selected to minimize the posterior error covariance. 

For the Kalman filter, the cost function was computed as the following mean-square error:
\begin{eqnarray}
    J_{t}= \mathbb{E}\left[(\epsilon _{x,t}^{+})^\top \epsilon _{x,t}^{+}\right]=\mathbb{E}\left[Tr(\epsilon _{x,t}^{+}(\epsilon _{x,t}^{+})^\top)\right] = Tr(P_{t}^{+})
    \label{eq: loss function}
\end{eqnarray}
where:
\begin{eqnarray}
    \begin{aligned}
\epsilon_{x,t}^{+}=x_{t}-\hat{x}_{t}^{+} & =x_{t}-\hat{x}_{t}^{-}-K_{t}\left(y_{t}-H_{t} \hat{x}_{t}^{-}\right) \\
& =\epsilon_{x, t}^{-}-K_{t}\left(H_{t} x_{t}+\eta _{t}-H_{t} \hat{x}_{t}^{-}\right) \\
& =\epsilon_{x, t}^{-}-K_{t} H_{t}\left(x_{t}-\hat{x}_{t}^{-}\right)-K_{t} \eta _{t} \\
& =\left(I-K_{t} H_{t}\right) \epsilon_{x, t}^{-}-K_{t} \eta _{t}
\end{aligned}
\label{eq: posterior estimate error}
\end{eqnarray}
denotes the posterior estimate error, 
\begin{eqnarray}
    P_{t}^{+} = \mathbb{E}\left[\epsilon _{x,t}^{+}(\epsilon _{x,t}^{+})^\top\right]
\end{eqnarray}
represents the posterior estimate error covariance. From Eq~(\ref{eq: posterior estimate error}), we obtained:
\begin{eqnarray}
    \begin{aligned}
P_{t}^{+} & =\mathbb{E}\left[\epsilon_{x, t}^{+} (\epsilon_{x, t}^{+})^\top\right] \\
& =\mathbb{E}\left[\left(\left(I-K_{t} H_{t}\right) \epsilon_{x, t}^{-}-K_{t} \eta _{t}\right)\left(\left(I-K_{t} H_{t}\right)\epsilon_{x, t}^{-}-K_{t}\eta _{t}\right)^\top\right] \\
& =\left(I-K_{t} H_{t}\right) \mathbb{E}\left[\epsilon_{x, t}^{-} (\epsilon_{x, t}^{-})^\top\right]\left(I-K_{t} H_{t}\right)^\top+K_{t} \mathbb{E}\left[\eta _{t} \eta _{t}^\top\right] K_{t}^\top \\
& =\left(I-K_{t} H_{t}\right) P_{t}^{-}\left(I-K_{t} H_{t}\right)^\top+K_{t} R_{t} K_{t}^\top
\end{aligned}
\label{eq: posterior estimate error covariance}
\end{eqnarray}
The first equation in Eq~(\ref{eq: analysis step}) was derived from setting the derivative of $J_{t}$ with respect to the Kalman gain matrix $K_{t}$ equal to zero:
\begin{eqnarray}
    \frac{\partial J_{t}}{\partial K_{t}}=\frac{\partial P_{t}^{+}}{\partial K_{t}}=2\left(I-K_{t} H_{t}\right) P_{t}^{-}\left(-H_{t}\right)^\top+2 K_{t} R_{t}
    \label{eq: derivative of Jt to Kt}=0
\end{eqnarray}

We considered the Kalman filter incorporating a quadratic regularization term. Specifically, a quadratic regularization term was added to the cost function of the Kalman filter:
\begin{eqnarray}
    \tilde{J_{t}}= J_{t} + \lambda \mathbb{E}\left[(\hat{x}_{t}^{+})^\top \Sigma \hat{x}_{t}^{+}\right]
    \label{eq:new loss function(1)}
\end{eqnarray}
where $\lambda$ denotes the penalty coefficient, $\Sigma$ is a semi-definite matrix. As before, to derive the updated formula for the Kalman gain matrix, we computed the derivative of $\tilde{J_{t}}$ with respect to $K_{t}$:
\begin{eqnarray}
    \frac{\partial \tilde{J_{t}}}{\partial K_{t}}=\frac{\partial J_{t}}{\partial K_{t}}+\lambda \frac{\partial}{\partial K_{t}} \mathbb{E}\left[(\hat{x}_{t}^{+})^\top \Sigma \hat{x}_{t}^{+}\right]
\end{eqnarray}
From simple derivation, it follows that:
\begin{eqnarray}
    \begin{aligned}
        \frac{\partial}{\partial K_{t}} \mathbb{E}\left[(\hat{x}_{t}^{+})^\top \Sigma \hat{x}_{t}^{+}\right] &= \mathbb{E}\left[\frac{\partial}{\partial K_{t}} ((\hat{x}_{t}^{+})^\top \Sigma \hat{x}_{t}^{+}) \right] \\
        &=\mathbb{E}\left[2\Sigma\hat{x}_{t}^{+}(y_{t}-H_{t}\hat{x}_{t}^{-})^\top\right]\\
        &=2\Sigma \mathbb{E}\left[(\hat{x}_{t}^{-} + K_{t}(y_{t}-H_{t}\hat{x}_{t}^{-}))(y_{t}-H_{t}\hat{x}_{t}^{-})^\top\right]\\
        &=2\Sigma (\mathbb{E}\left[\hat{x}_{t}^{-}(y_{t}-H_{t}\hat{x}_{t}^{-})^\top\right]+K_{t}\mathbb{E}\left[H_{t}\epsilon_{x,t}^{-}(\epsilon_{x,t}^{-})^\top H_{t}^\top \right]+K_{t}R_{t})\\
        &=2\Sigma (\mathbb{E}\left[\hat{x}_{t}^{-}(y_{t}-H_{t}\hat{x}_{t}^{-})^\top\right]+K_{t}(H_{t}P_{t}^{-}H_{t}^\top+R_{t}))
    \end{aligned}
    \label{eq: derivative of regularization term}
\end{eqnarray}
By combining Eq~(\ref{eq: derivative of Jt to Kt}) and Eq~(\ref{eq: derivative of regularization term}), the derivative of $\tilde{J_{t}}$ with respect to $K_{t}$ was given by:
\begin{eqnarray}
    \frac{\partial \tilde{J_{t}}}{\partial K_{t}}=2(I+\lambda \Sigma)K_{t}(H_{t}P_{t}^{-}H_{t}^\top+R_{t})-2P_{t}^{-}H_{t}^\top+2\lambda \Sigma \mathbb{E}\left[\hat{x}_{t}^{-}(y_{t}-H_{t}\hat{x}_{t}^{-})^\top
    \right]
\end{eqnarray}
Then, setting the derivative equal to zero yielded the updated formula for the Kalman gain matrix:
\begin{eqnarray}
    K_{t} = (I + \lambda  \Sigma)^{-1}\left[P_{t}^{-} H_{t}^\top - \lambda \Sigma \mathbb{E}\left[\hat{x}_{t}^{-}(y_{t}-H_{t}\hat{x}_{t}^{-})^\top\right]\right](H_{t} P_{t}^{-} H_{t}^\top+R_{t})^{-1}
    \label{eq: si new Kalman gain matrix}
\end{eqnarray}

\subsubsection{Regularized-HDA algorithm}

As described in the main text, to resolve the ill-posed inverse problem (where hyperparameter dimensions exceed electrode count), we derived Regularized-HDA by incorporating a geometry-constrained regularization term:
\begin{equation}
L_{p}=\sum_{i=1}^{m} \sum_{j \in \mathcal{A}_{i} }|h_{j} - \bar{h}_{i}|^{2}  \label{eq:si Laplace regularization}
\end{equation}
where:
\begin{equation}
    \bar{h}_{i} = \frac{\sum_{j\in \mathcal{A}_{i}} h_{j}}{\# \mathcal{A}_{i}}
\end{equation}
denotes the mean of the hyperparameters of voxels $\mathcal{A}_{i}$. We expressed the penalty term $L_{p}$ using the system's state variables $\mathbf{x}$, rewriting Eq ~\ref{eq:si Laplace regularization} as:
\begin{eqnarray}
    L_{p} = \mathbb{E}\left[\left \| T_{1}\mathbf{x}-T_{2}T_{1}\mathbf{x} \right \|_{2}^{2}\right]
    \label{eq:laplace penalty term}
\end{eqnarray}
where $T_{1}$ projects $\mathbf{x}$ to hyperparameters:
\begin{eqnarray*}
    \mathbf{h} = T_{1}\mathbf{x}=[\mathbf{h_{1}}, \mathbf{h_{2}}, \cdots, \mathbf{h_{m}}]
\end{eqnarray*}
with $\mathbf{h_{i}} = [h_{j}]_{j \in \mathcal{A}_{i}}$, $T_{2}$ projects hyperparameters $\mathbf{h}$ to means $\mathbf{\bar{h}}$:
\begin{eqnarray*}
    \mathbf{\bar{h}}=T_{2}T_{1}\mathbf{x}=T_{2}\mathbf{h}=[\bar{h}_{1}\mathbf{e_{1}}, \bar{h}_{2}\mathbf{e_{2}}, \cdots, \bar{h}_{m}\mathbf{e_{m}}]
\end{eqnarray*}
where $\mathbf{e_{i}} \in \mathbb{R}^{\# \mathcal{A}_{i} \times 1} $ is a unit vector. We added the Laplacian regularization term Eq~(\ref{eq:laplace penalty term}) to the Kalman filter's cost function. As derived in the last section, the updated formula for the Kalman gain matrix was given by:
\begin{eqnarray}
    K = (I + \lambda  \mathcal{L})^{-1}\left[\hat{P} H^\top - \lambda \mathcal{L} \mathbb{E}\left[\hat{x}(y-H\hat{x})^\top\right]\right](H\hat{P} H^\top+\Gamma)^{-1}
    \label{eq: new Kalman gain matrix(2)}
\end{eqnarray}
where $\mathcal{L}=(T_{1}-T_{2}T_{1})^\top(T_{1}-T_{2}T_{1})$, referred to as the penalty matrix, $H$ is the matrix mapping the state variables to the observations, $\hat{x}$ represents the prior estimate of the state variables, $\hat{P}$ is the prior estimate error covariance matrix of state variables, $y$ denotes the observations at a given time point, $\Gamma$ represents the observation noise covariance matrix, $I$ is the identity matrix. 

We presented the pseudo-code for our Regularized-HDA as follows:
\begin{algorithm}
\SetAlgoNoLine
\caption{Regularized-HDA}\label{algorithm 1}
\KwIn{Ensemble members number $N$, Initial distribution parameters $\left [ \mu_{0}, P_{0} \right ]$, Evolution function $F_{0:T}(\cdot)$, Hyperparameter random walk covariance matrix $ {\textstyle \sum_{h}}$, Model noise covariance matrix $ {\textstyle \sum_{x}}$, Observation matrix $\left \{ H_{i} \right \}_{i=1}^{m}$, Observation noise covariance matrix $\left \{ \Gamma_{i} \right \}_{i=1}^{m}$, Observation sequence of all sensors $\left \{ y_{i, 0:T} \right \}_{i=1}^{m}$, Fusion coefficient $\gamma$, Penalty matrix $\mathcal{L}$, Penalty coefficient $\lambda$}
\KwOut{Local estimation of all sensors $\left \{ x_{i,t}^{(n)} | i=1,\cdots,m; t=0:T\right \}_{n=1}^{N}$}

\textbf{draw} $x_{0}^{(n)}\sim \mathcal{N}(\mu_{0}, P_{0}), \quad \theta_{0}^{(n)}\sim \mathbb{P}_{\varphi}(\cdot|\hat{h}_{0}^{(n)}), \quad \hat{h}_{0}^{(n)}=h_{0}^{(n)}, \quad \forall n=1:N $\;
\For{$t=1:T$}{
        \For{$i=1:m$}{
        \textbf{draw} $\hat{h}_{i,t}^{(n)}\sim \mathcal{N}(h_{i,t-1}^{(n)},  {\textstyle \sum_{h}}), \quad \forall n=1:N $\;
        $\theta_{i,t}^{(n)}=\mathcal{F}_{\hat{h}_{i,t}^{(n)}}^{-1} \circ \mathcal{F}_{h_{i,t-1}^{(n)}} (\theta_{i,t-1}^{(n)}), \quad \forall n=1:N$\;
        \textbf{draw} $\hat{x}_{i,t}^{(n)}\sim\mathcal{N}(F_{t-1}(x_{i,t-1}^{(n)}, \theta_{i,t}^{(n)}),  {\textstyle \sum_{x}}), \quad \forall n=1:N$\;
        $\hat{\mu}_{i,t}=\frac{1}{N}\sum_{n=1}^{N}\hat{x}_{i,t}^{(n)}$\;
        $\hat{P}_{i,t}=\frac{1}{N-1}\sum_{n=1}^{N}(\hat{x}_{i,t}^{(n)}-\hat{\mu}_{i,t})(\hat{x}_{i,t}^{(n)}-\hat{\mu}_{i,t})^\top$\;
        \For {any sensor $j\in\mathcal{N}(i)$}{
        $S_{j, t}=H_{j}\hat{P}_{i,t}H_{j}^\top+\Gamma_{j}$\;
        \textbf{draw} $\bigtriangleup y_{j,t}^{(n)}\sim\mathcal{N}(0, \Gamma_{j}), \quad \forall n=1:N$\;
        $\delta_{j,t}^{(n)}=y_{j,t}+\bigtriangleup y_{j,t}^{(n)}-H_{j}\hat{x}_{i,t}^{(n)}, \quad \forall n=1:N$\;
        $K_{j, t} = (I + \lambda  \mathcal{L})^{-1}\left[\hat{P}_{i, t} H_{j}^\top - \lambda \mathcal{L} \frac{1}{N} \sum_{n=1}^{N}\hat{x}_{i, t}^{(n)}(\delta_{j,t}^{(n)})^\top\right]S_{j, t}^{-1}$\;
        $\hat{x}_{i,
        t}^{(n)}=\hat{x}_{i,t}^{(n)}+K_{j,t}\delta_{j,t}^{(n)}, \quad \forall n=1:N$\; 
        }
        $x_{i,t}^{(n)}=\sum_{l \in \mathcal{N}(i)} \mathcal{C}_{l, i}(\gamma) \hat{x}_{l, t}^{(n)} ; \quad \forall n=1:N$\;
        }
        \textbf{return} $\left \{ x_{i,t}^{(n)}| i=1,\cdots,m \right \}_{n=1}^{N}$\;
		
	}
	\textbf{return} $\left \{ x_{i,t}^{(n)}|i=1,\cdots,m;t=0,\cdots,T \right \}_{n=1}^{N}$.
\end{algorithm}

In Algorithm ~\ref{algorithm 1}, $\mathbb{P}_{\varphi}(\cdot|\hat{h}_{0}^{(n)})$ denotes a Gamma distribution with shape parameter $\alpha=5$ and inverse scale parameter $\beta= 5 / \hat{h}_{0}^{(n)}$. $\mathcal{F}_{h}$ and $\mathcal{F}_{h}^{-1}$ represent the cumulative density function (cdf) and inverse cdf of this Gamma distribution(with hyperparameter $h$), respectively. For sensor $i$, $\mathcal{N}(i)$ represents its neighborhood, while $\mathcal{C}_{l,i}(\gamma)$ weights neighboring sensor estimates via fusion coefficient $\gamma$. Here, we considered a fully connected sensor network, and $\mathcal{C}_{l, i}$ was defined as:
\begin{align}
    \mathcal{C}_{l, i} &= diag\left \{ c_{1;l,i}I_{n_{1}}, \cdots, c_{m;l,i}I_{n_{m}} \right \} ; \\
    c_{r;l,i}&=\left\{\begin{matrix}
 \gamma,  & if \quad  r=l\\
 \frac{1-\gamma }{m -1 },  & else
\end{matrix}\right.
\end{align}
where $I_{n_{i}} \in \mathbb{R}^{n_{i} \times n_{i}}$ was the unity matrix, $n_{i}$ was the dimension of sub-state $x_{i}$.

\subsection{EEGNet-LSTM}

Building upon the regression model architecture in Table 2 of ~\cite{ALawhern2018}, we presented our network structure in Table~\ref{EEGNet-table}.
\begin{table}
  \small 
  \caption{EEGNet-LSTM architecture.}
  \label{EEGNet-table}
  \centering
  \begin{threeparttable}
  \begin{tabular}{llllllll}
    \toprule
    Block     & Layer     & \# Filters & Size & \# Params & Output shape & Activation & Options  \\
    \midrule
    \multirow{9}{*}{1} & Input & - & - & -  & (B, C,T) & - & - \\
      & Reshape & - & - & - & (B, 1,C,T) & -  & dim=1 \\
      & \multirow{2}{*}{Conv2D} & \multirow{2}{*}{$F_{1}$}  & \multirow{2}{*}{(1, $K_{1}$)} & \multirow{2}{*}{$1 \times F_{1} \times K_{1}$} & \multirow{2}{*}{(B, $F_{1}$, C, $T$)} & \multirow{2}{*}{ELU} & padding= \\
      & & & & & & & (0, $K_{1} // 2$) \\
      & BatchNorm2D & $F_{1}$  & - & $2 \times F_{1}$ & (B, $F_{1}$, C, $T$) & - & - \\
      & Depthwise Conv2D  & $F_{1} \times D$ & $(C,1)$ & $F_{1} \times 1 \times C$ & (B, $F_{1} \times D$, 1, $T$) & ELU & groups=$F_{1}$ \\
      & BatchNorm2D & $F_{1} \times D$ & - & $2 \times F_{1} \times D$ & (B, $F_{1} \times D$, 1, $T$) & - & - \\
      & AvgPool2D & - & (1, $P_{1}$) & - & (B, $F_{1} \times D$, 1, $T / P_{1}$) & - & stride=$P_{1}$ \\
      & Dropout & - & - & - & (B, $F_{1} \times D$, 1, $T / P_{1}$) & - & p=$p_{1}$ \\
      \midrule
    \multirow{7}{*}{2} & \multirow{2}{*}{Separable Conv2D} & \multirow{2}{*}{$F_{2}$} & \multirow{2}{*}{(1, $K_{2}$)} & \multirow{2}{*}{$F_{2} \times F_{2} \times 1 \times K_{2}$} & \multirow{2}{*}{(B, $F_{2}$, 1, $T / P_{1}$)} & \multirow{2}{*}{ELU} & padding= \\
    & & & & & & & (0, $K_{2} // 2$)  \\
      & BatchNorm2D & $F_{2}$ & - & $2 \times F_{2}$ & (B, $F_{2}$, 1, $T / P_{1}$) & - & - \\
      & AvgPool2D & - & (1, $P_{2}$) & - & (B, $F_{2}$, 1, $T / P_{1} / P_{2}$) & - & stride=$P_{2}$ \\
      & Dropout & - & - & - & (B, $F_{2}$, 1, $T / P_{1} / P_{2}$) & - & p=$p_{2}$ \\
      & Squeeze & - & - & - & (B, $F_{2}$, $T / P_{1} / P_{2}$) & - & dim=2 \\
      & Reshape & - & - & - & (B, $T / P_{1} / P_{2}$, $F_{2}$) & - & - \\
      \midrule
    \multirow{5}{*}{3} & \multirow{2}{*}{LSTM(2-layer)} & \multirow{2}{*}{H} & \multirow{2}{*}{-} & $ 4 \times(F_{2}$ & \multirow{2}{*}{(B, $T / P_{1} / P_{2}$, H)} & \multirow{2}{*}{-} & \multirow{2}{*}{-} \\
    & & & & $+H)\times H \times 2$ & & & \\
      & Select last step & - & - & - & (B, H) & - & out[:, -1, :]\\
      & Fully connected & 1 & - & $H \times 1$ & (B, 1) & Linear & - \\
      & Regression Output & - & - & - & (B,) & - & - \\
    \bottomrule 
  \end{tabular}
  \end{threeparttable}
\end{table}
The parameter definitions and values are provided in Table ~\ref{tab: Parameter definitions and values}.
\begin{table}[!ht]
  \centering
  \caption{{\bf Parameter definitions and values in the EEGNet-LSTM architecture}}
  \label{tab: Parameter definitions and values}
  \begin{tabular}{ccccccccccc}
    \toprule
    {\bf Symbol} & {\bf Value} & {\bf Definition}   \\
    \toprule
    $C$ & $ 63 $ & Number of channels \\
    \midrule
    $T$ & $200$ & Number of time points per EEG patch \\
    \midrule
    $F_{1} / F_{2}$ & $8 / 16$ & Number of filters in the first / Separable Conv2D layer \\
    \midrule
    $K_{1}/K_{2}$ & $64/16$ & Kernel size in the first/Saparable Conv2D layer \\
    \midrule
    $D$ & $2$ & Depth multiplier for Depthwise Conv2D \\
    \midrule
    $P_{1}/P_{2}$ & $4/8$ & Pooling size in first/second block\\
    \midrule
    $H$ & $64$ & Hidden size of LSTM layers \\
    \midrule
    $p_{1} / p_{2}$ & $0.5/0.5$ & Dropout rate after first/second block \\
    \bottomrule
  \end{tabular}
\end{table}

\subsection{Hardware and software environment}

The C++ program, developed using the g++ compiler (version 7.3.1), incorporates MPI library functions and utilizes the ROCm-4.5 HIPCC compiler for GPU acceleration. The Python program for data assimilation was implemented using Python version 3.7.11. Both Python and C++ programs were executed in a Linux environment.

The cluster consists of 270 computing nodes, each equipped with a single 32-core processor operating at 2 GHz and 128 GB of DRAM. Additionally, each node is outfitted with 4 GPUs, each running at 1.10 GHz and featuring 16GB of HBM2 memory operating at 800 MHz, with a memory bandwidth of 1TB/s. Communication between GPUs within a node is facilitated via shared memory, while inter-node communication occurs over a 200 Gbps Full Duplex Infiniband network.

\subsection{Supplementary results}

Parameters for Algorithm~\ref{algorithm 1} are specified in Table ~\ref{tab: Parameter values} for resting-state data assimilation, with the sole difference being the random walk step size ($\sigma_{h} = 2$) during task-state assimilation.
\begin{table}[!ht]
  \centering
  \caption{{\bf Parameter values in Algorithm ~\ref{algorithm 1} for resting-state assimilation}}
  \label{tab: Parameter values}
  \begin{tabular}{ccccccccccc}
    \toprule
    {\bf Symbol} & {\bf Value} & {\bf Definition}   \\
    \toprule
    $\sigma_{h}$ & $ 1.1 $ & Step size of random walk \\
    \midrule
    $\sigma_{o}$ & $10^{-5}$ & Standard deviation of observation noise \\
    \midrule
    $N$ & $80$ & Ensemble number \\
    \midrule
    $\gamma$ & $0.5$ & Fusion coefficient \\
    \midrule
    $\lambda$ & $0.001$ & Penalty coefficient \\
    \bottomrule
  \end{tabular}
\end{table}

\begin{figure}[t!]
  \centering
  \includegraphics[width=5.5in]{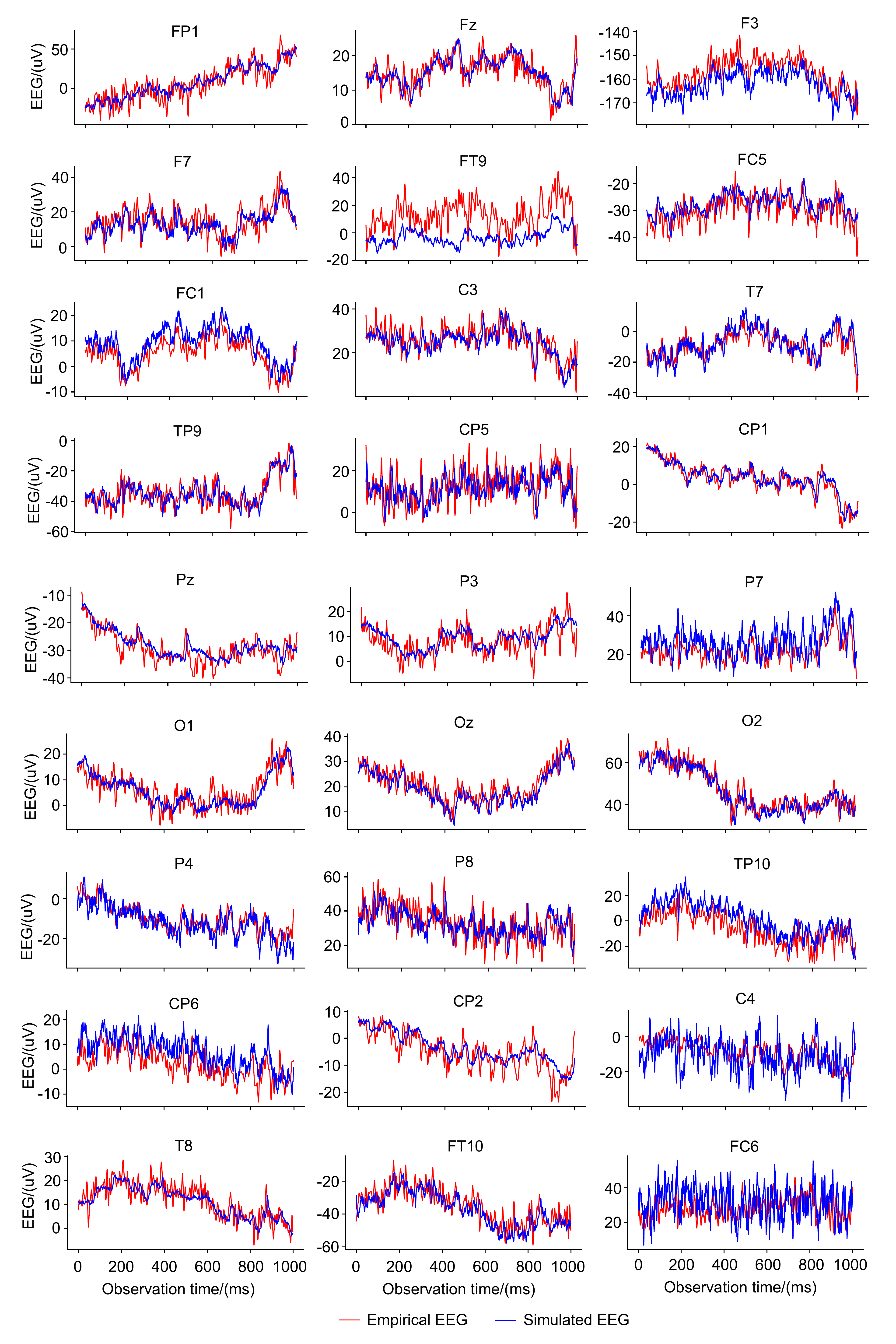}
  \caption{ {\bf Resting-state EEG-DTB simulation(from FP1 to FC6 electrodes).} Comparison of simulated (blue) and empirical (red) resting-state EEG signals.}
  \label{fig:si fig2}
\end{figure}

\begin{figure}[t!]
  \centering
  \includegraphics[width=5.5in]{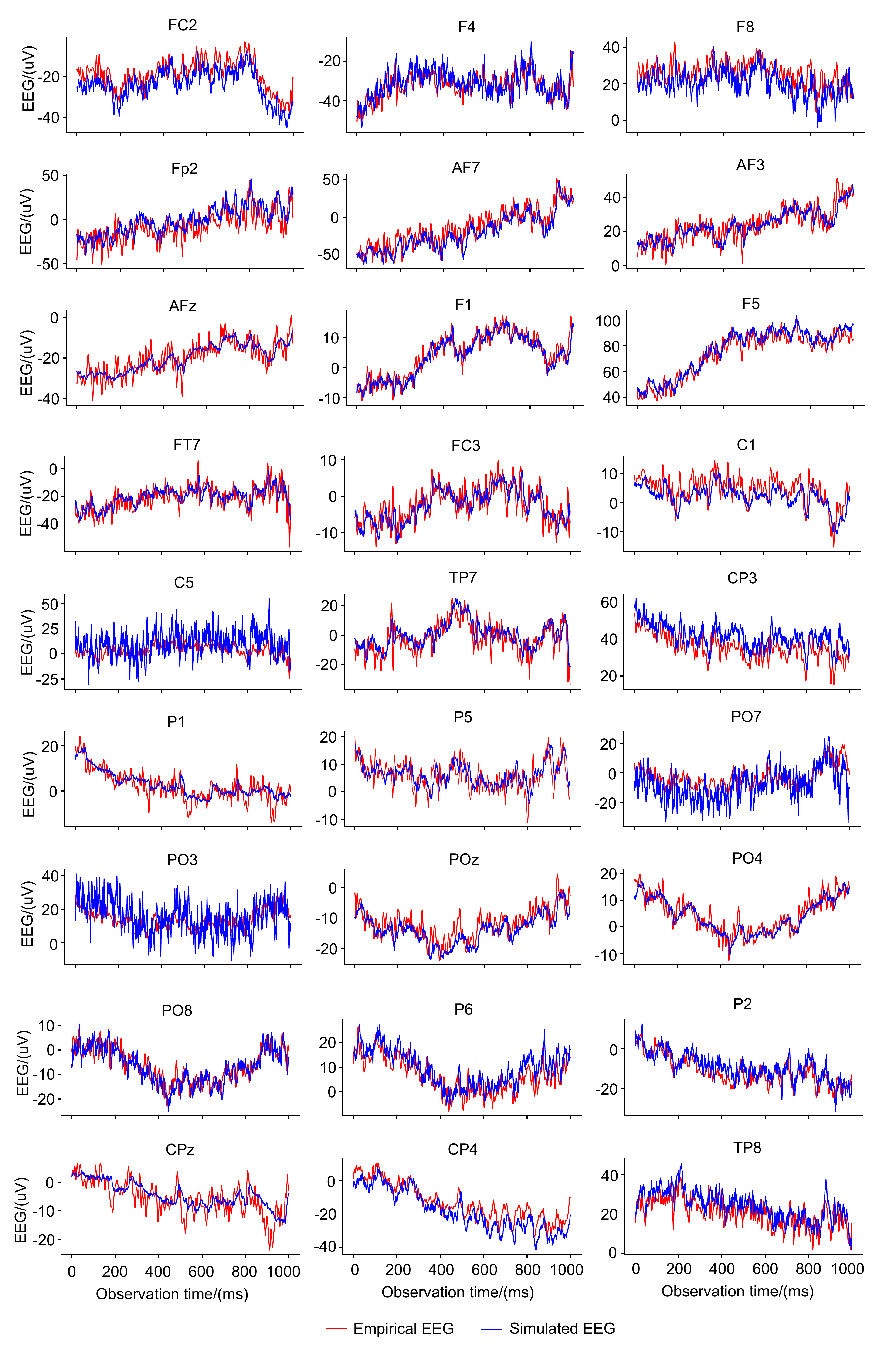}
  \caption{ {\bf Resting-state EEG-DTB simulation(from FC2 to TP8 electrodes).} Comparison of simulated (blue) and empirical (red) resting-state EEG signals.}
  \label{fig:si fig3}
\end{figure}

\begin{figure}[t!]
  \centering
  \includegraphics[width=5in]{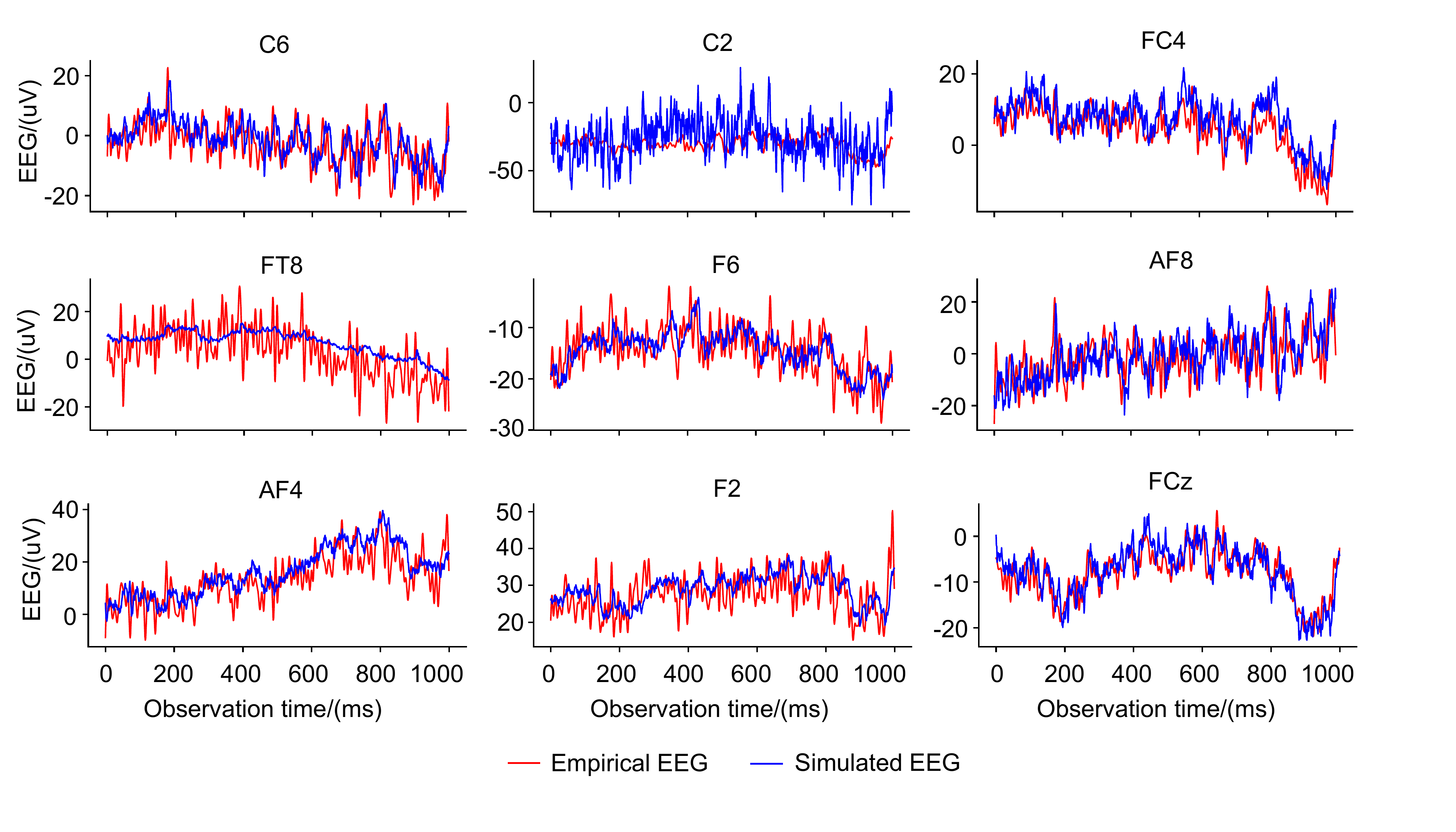}
  \caption{ {\bf Resting-state EEG-DTB simulation(from C6 to FCz electrodes).} Comparison of simulated (blue) and empirical (red) resting-state EEG signals.}
  \label{fig:si fig1}
\end{figure}

\begin{figure}[t!]
  \centering
  \includegraphics[width=5.5in]{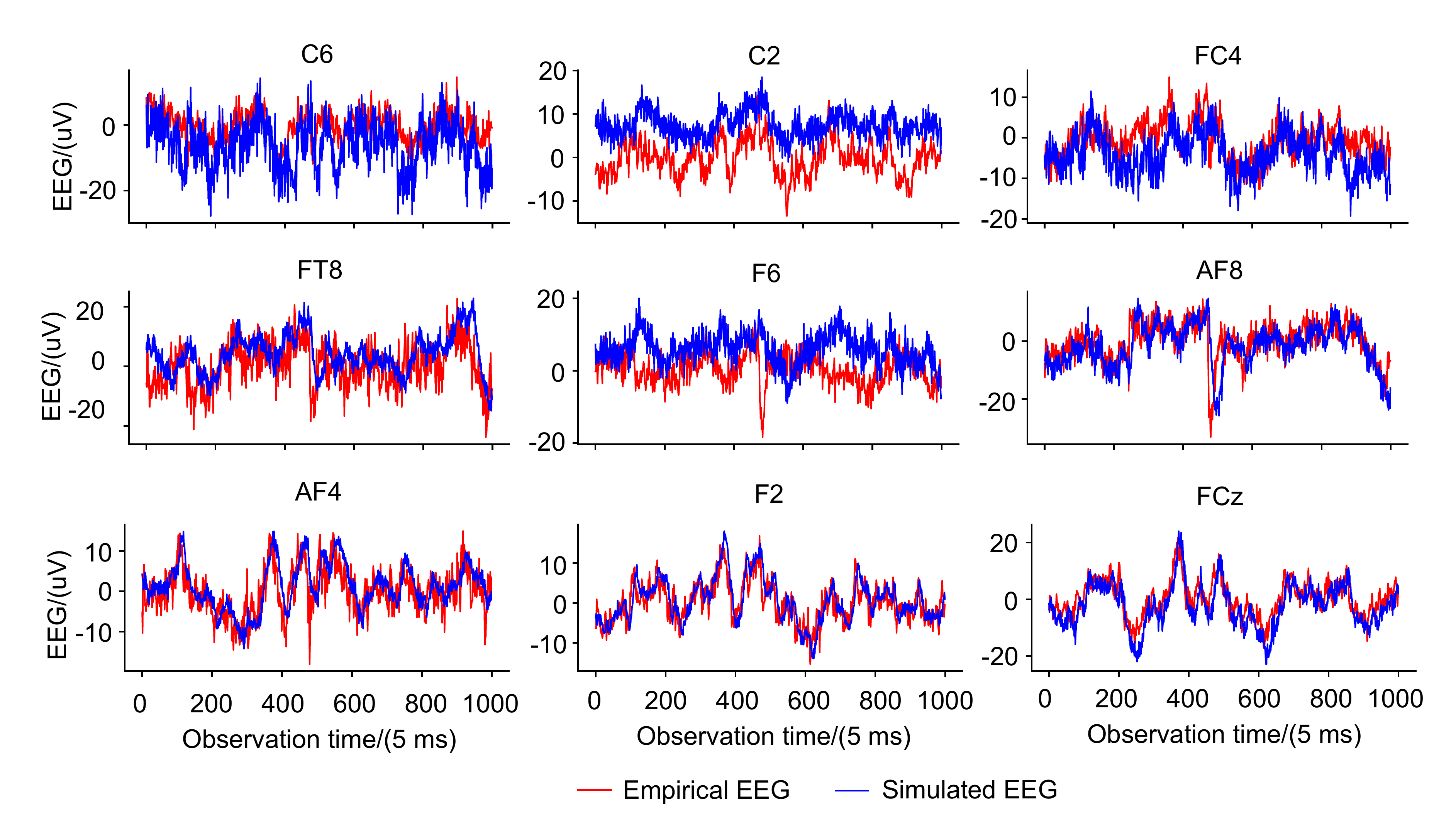}
  \caption{ {\bf Task-state EEG-DTB simulation(from C6 to FCz electrodes).} Comparison of simulated (blue) and empirical (red) task-state EEG signals.}
  \label{fig:si fig6}
\end{figure}

\begin{figure}[t!]
  \centering
  \includegraphics[width=5.5in]{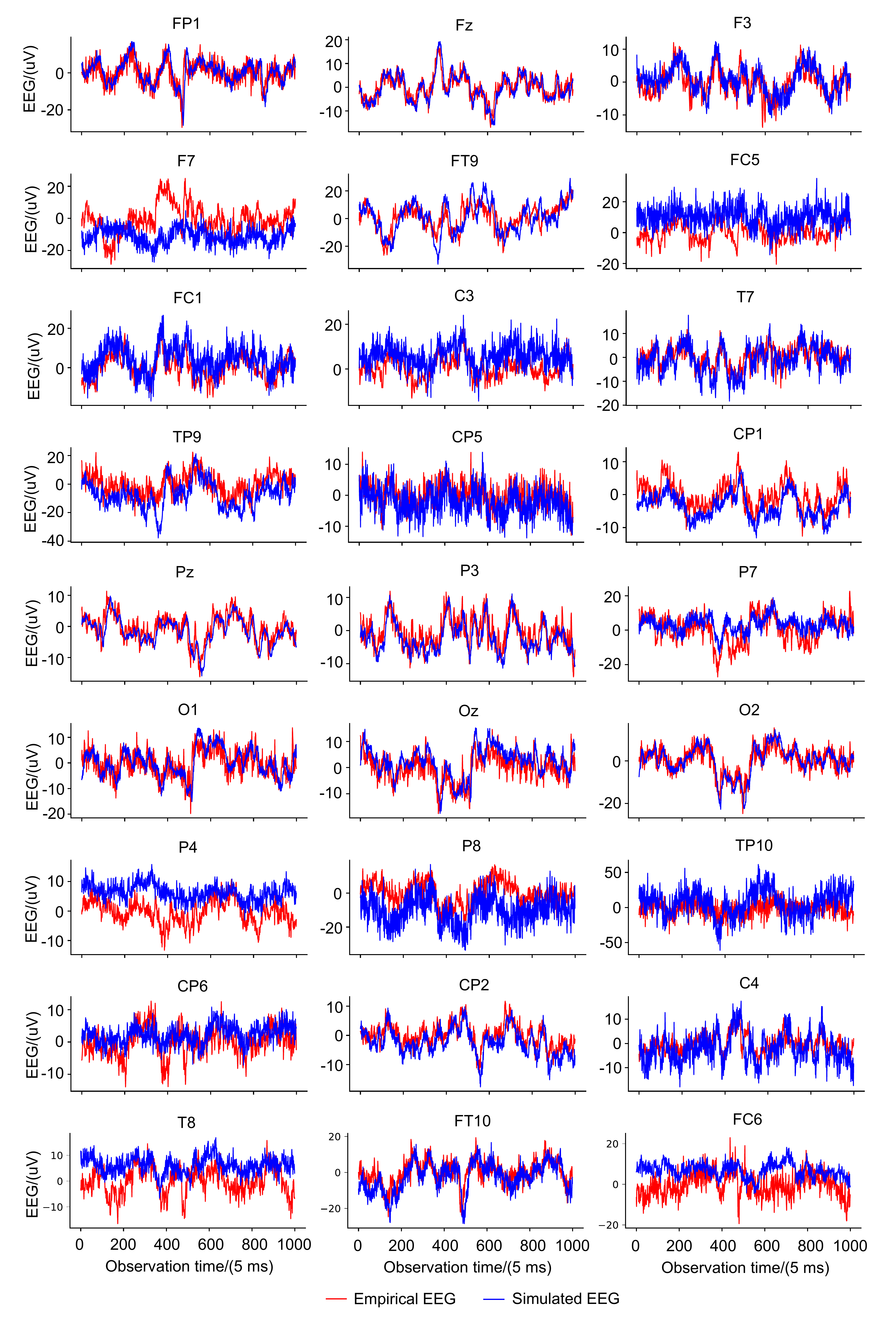}
  \caption{ {\bf Task-state EEG-DTB simulation(from FP1 to FC6 electrodes).} Comparison of simulated (blue) and empirical (red) task-state EEG signals.}
  \label{fig:si fig4}
\end{figure}

\begin{figure}[t!]
  \centering
  \includegraphics[width=5.5in]{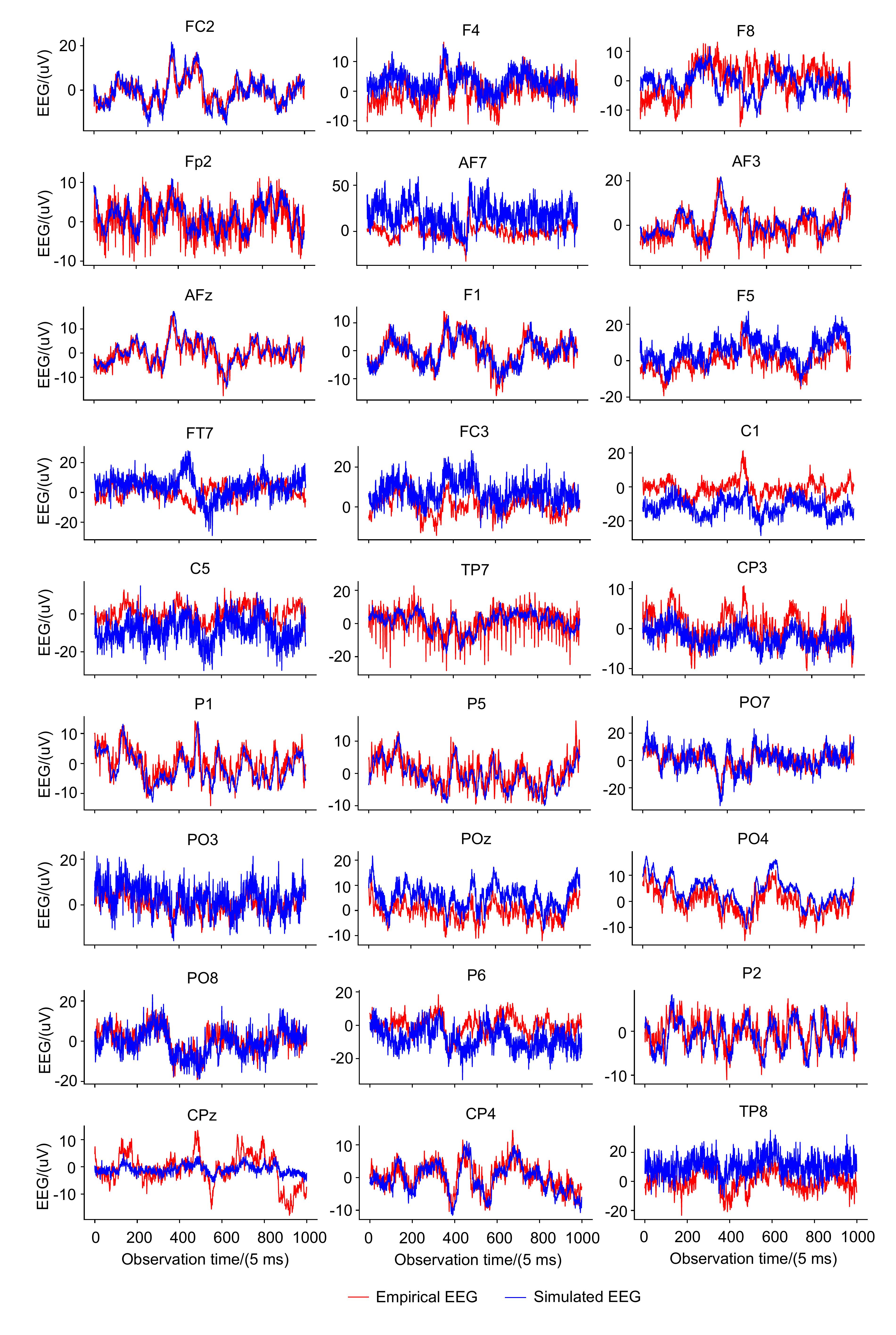}
  \caption{ {\bf Task-state EEG-DTB simulation(from FC2 to TP8 electrodes).} Comparison of simulated (blue) and empirical (red) task-state EEG signals.}
  \label{fig:si fig5}
\end{figure}

Fig ~\ref{fig:si fig2}-\ref{fig:si fig1} and Fig ~\ref{fig:si fig6}-\ref{fig:si fig5} display simulated (blue) versus empirical (red) EEG signals for all electrodes during resting-state and task-state conditions, respectively.

The simulated EEG data and decoded steering angles are available at: \href{https://anonymous.4open.science/r/test-D10B/}{https://anonymous.4open.science/r/test-D10B/}.

\FloatBarrier

\FloatBarrier

\FloatBarrier

\FloatBarrier

\FloatBarrier

\FloatBarrier


\bibliographystyle{unsrt}

\begin{thebibliography}{10} 

\bibitem{stein1965theoretical}
Stein RB.
\newblock A theoretical analysis of neuronal variability.
\newblock Biophysical journal. 1965 Mar;5(2):173--194.

\bibitem{burkitt2006review}
Burkitt AN.
\newblock A review of the integrate-and-fire neuron model: I. Homogeneous synaptic input.
\newblock Biological cybernetics. 2006 Jul;95(1):1--19.

\bibitem{hodgkin1952quantitative}
Hodgkin AL, Huxley AF.
\newblock A quantitative description of membrane current and its application to conduction and excitation in nerve.
\newblock The Journal of physiology. 1952 Aug;117(4):500-544.

\bibitem{wilson1972excitatory}
Wilson HR, Cowan JD.
\newblock Excitatory and inhibitory interactions in localized populations of model neurons.
\newblock Biophysical journal. 1972;12(1):1--24.

\bibitem{friston2014nodes}
Friston KJ, Kahan J, Razi A, Stephan KE, Sporns O.
\newblock On nodes and modes in resting state fMRI.
\newblock NeuroImage. 2014 Oct 1;99:533--547.

\bibitem{Pogossian1975}
Ter-Pogossian, M. M., Phelps, M. E., Hoffman, E. J., Mullani, N. A.
\newblock A positron-emission transaxial tomograph for nuclear imaging (PETT).
\newblock Radiology. 1975;114(1):89-98. 

\bibitem{berger1931elektrenkephalogramm}
Berger H.
\newblock {\"U}ber das elektrenkephalogramm des menschen.
\newblock Archiv f{\"u}r Psychiatrie und Nervenkrankheiten. 1929;87:527--570.

\bibitem{michel2012towards}
Michel CM, Murray MM.
\newblock Towards the utilization of EEG as a brain imaging tool.
\newblock Neuroimage. 2012;61(2):371--385.

\bibitem{boto2018moving}
Boto E, Holmes N, Leggett J, Roberts G, Shah V, Meyer SS, et~al.
\newblock Moving magnetoencephalography towards real-world applications with a wearable system.
\newblock Nature. 2018;555(7698):657--661.

\bibitem{eibern1999four}
Eibern H, Schmidt H.
\newblock A four-dimensional variational chemistry data assimilation scheme for Eulerian chemistry transport modeling.
\newblock Journal of Geophysical Research: Atmospheres. 1999;104(D15):18583--18598.

\bibitem{Lopes2011}
Lopes, H. F., Tsay, R. S.
\newblock Particle filters and Bayesian inference in financial
econometrics.
\newblock Journal of Forecasting. 2011;30:168–209. 

\bibitem{politi2016comparing}
Politi N, Feng J, Lu W.
\newblock Comparing data assimilation filters for parameter estimation in a neuron model.
\newblock In: 2016 international joint conference on neural networks (IJCNN). IEEE; 2016. pp. 4767--4774.

\bibitem{zhang2024framework}
Zhang W, Chen B, Feng J, Lu W.
\newblock On a framework of data assimilation for hyperparameter estimation of spiking neuronal networks.
\newblock Neural Networks. 2024 Mar;171:293--307.

\bibitem{escuain2018extracranial}
Escuain-Poole L, Garcia-Ojalvo J, Pons AJ.
\newblock Extracranial estimation of neural mass model parameters using the unscented Kalman filter.
\newblock Frontiers in Applied Mathematics and Statistics. 2018;4:46.

\bibitem{yokoyama2023data}
Yokoyama H, Kitajo K.
\newblock A data assimilation method to track excitation-inhibition balance change using scalp EEG.
\newblock Communications Engineering. 2023;2(1):92.

\bibitem{jansen1995electroencephalogram}
Jansen BH, Rit VG.
\newblock Electroencephalogram and visual evoked potential generation in a mathematical model of coupled cortical columns.
\newblock Biological cybernetics. 1995;73(4):357--366.

\bibitem{Ramezani2025}
Ramezani, M., Ren, Y., Cubukcu, E. et al.
\newblock Innovating beyond electrophysiology through multimodal neural interfaces.
\newblock Nat Rev Electr Eng. 2025;2:42–57.

\bibitem{zhang2023deep}
Zhang W, Lu W.
\newblock Deep diffusion Kalman filter combining large-scale neuronal networks simulation with multimodal neuroimaging data.
\newblock Mathematics. 2023;11(12):2716.

\bibitem{2023EEGNetLSTM}
de Oliveira IH, Rodrigues AC.
\newblock Empirical comparison of deep learning methods for EEG decoding. 
\newblock Front. Neurosci. 2023;16:1003984.

\bibitem{Lawhern2018}
Lawhern, V. J., Solon, A. J., Waytowich, N. R., Gordon, S. M., Hung, C. P., Lance, B. J.
\newblock EEGNet: a compact convolutional neural network for EEG-based brain-computer interfaces.
\newblock Journal of neural engineering. 2018;15(5):056013.

\bibitem{gramfort2013meg}
Gramfort A, Luessi M, Larson E, Engemann DA, Strohmeier D, Brodbeck C, et~al.
\newblock MEG and EEG data analysis with MNE-Python.
\newblock Frontiers in Neuroinformatics. 2013 Dec 26;7:267.

\bibitem{hyvarinen1999fast}
Hyvarinen A.
\newblock Fast and robust fixed-point algorithms for independent component analysis.
\newblock IEEE transactions on Neural Networks. 1999;10(3):626--634.

\bibitem{lu2024imitating}
Lu W, Zeng L, Wang J, Xiang S, Qi Y, Zheng Q, et~al.
\newblock Imitating and exploring the human brain's resting and task-performing states via brain computing: scaling and architecture.
\newblock National Science Review. 2024 Mar 1;11(5):nwae080.

\bibitem{brunel2001effects}
Brunel N, Wang XJ.
\newblock Effects of neuromodulation in a cortical network model of object working memory dominated by recurrent inhibition.
\newblock Journal of computational neuroscience. 2001;11:63--85.

\bibitem{yang2016dendritic}
Yang GR, Murray JD, Wang XJ.
\newblock A dendritic disinhibitory circuit mechanism for pathway-specific gating.
\newblock Nature communications. 2016 Sep 20;7(1):12815.

\bibitem{lu2022human}
Lu W, Zheng Q, Xu N, Feng J, Consortium D.
\newblock The human digital twin brain in the resting state and in action.
\newblock arXiv:2211.15963.[Preprint]. 2022 nov.

\bibitem{huang2022extended}
Huang CC, Rolls ET, Feng J, Lin CP.
\newblock An extended Human Connectome Project multimodal parcellation atlas of the human cortex and subcortical areas.
\newblock Brain Structure and Function. 2022 Apr;227(3):763--778.

\bibitem{binzegger2004quantitative}
Binzegger T, Douglas RJ, Martin KA.
\newblock A quantitative map of the circuit of cat primary visual cortex.
\newblock Journal of Neuroscience. 2004 Sep 29;24(39):8441--8453.

\bibitem{PMID:23093925}
Du J, Vegh V, Reutens DC.
\newblock The laminar cortex model: a new continuum cortex model incorporating laminar architecture.
\newblock PLoS computational biology. 2012;8(10):e1002733.

\bibitem{Mosher1999}
Mosher, J. C., Leahy, R. M., Lewis, P. S.
\newblock EEG and MEG: forward solutions for inverse methods.
\newblock IEEE transactions on bio-medical engineering. 1999;46(3):245–259.

\bibitem{Tripp1983}
J.~H. Tripp.
\newblock Physical Concepts and Mathematical Models.
\newblock Boston, MA: Springer US. 1983;101--139.  

\bibitem{Gramfort2010}
Gramfort A, Papadopoulo T, Olivi E, Clerc M.
\newblock OpenMEEG: opensource software for quasistatic bioelectromagnetics. 
\newblock BioMedical Engineering OnLine. 2010;45:9.

\bibitem{kuhn1955hungarian}
Kuhn HW.
\newblock The Hungarian method for the assignment problem.
\newblock Naval research logistics quarterly. 1955;2(1-2):83--97.




\end{thebibliography}

\begin{thebibliography}{10} 

\bibitem{Abrunel2001effects}
Brunel N, Wang XJ.
\newblock Effects of neuromodulation in a cortical network model of object working memory dominated by recurrent inhibition.
\newblock Journal of computational neuroscience. 2001;11:63--85.

\bibitem{Ayang2016dendritic}
Yang GR, Murray JD, Wang XJ.
\newblock A dendritic disinhibitory circuit mechanism for pathway-specific gating.
\newblock Nature communications. 2016 Sep 20;7(1):12815.

\bibitem{Alu2022human}
Lu W, Zheng Q, Xu N, Feng J, Consortium D.
\newblock The human digital twin brain in the resting state and in action.
\newblock arXiv:2211.15963.[Preprint]. 2022 nov.

\bibitem{Abinzegger2004quantitative}
Binzegger T, Douglas RJ, Martin KA.
\newblock A quantitative map of the circuit of cat primary visual cortex.
\newblock Journal of Neuroscience. 2004 Sep 29;24(39):8441--8453.

\bibitem{APMID:23093925}
Du J, Vegh V, Reutens DC.
\newblock The laminar cortex model: a new continuum cortex model incorporating laminar architecture.
\newblock PLoS computational biology. 2012;8(10):e1002733.

\bibitem{Afalchier2002anatomical}
Falchier A, Clavagnier S, Barone P, Kennedy H.
\newblock Anatomical evidence of multimodal integration in primate striate cortex.
\newblock Journal of Neuroscience. 2002 Jul 1;22(13):5749--5759.

\bibitem{Atournier2019mrtrix3}
Tournier JD, Smith R, Raffelt D, Tabbara R, Dhollander T, Pietsch M, et~al.
\newblock MRtrix3: A fast, flexible and open software framework for medical image processing and visualisation.
\newblock Neuroimage. 2019;202:116137.

\bibitem{Averaart2016denoising}
Veraart J, Novikov DS, Christiaens D, Ades-Aron B, Sijbers J, Fieremans E.
\newblock Denoising of diffusion MRI using random matrix theory.
\newblock Neuroimage. 2016;142:394--406.

\bibitem{Averaart2016diffusion}
Veraart J, Fieremans E, Novikov DS.
\newblock Diffusion MRI noise mapping using random matrix theory.
\newblock Magnetic resonance in medicine. 2016;76(5):1582--1593.

\bibitem{Acordero2019complex}
Cordero-Grande L, Christiaens D, Hutter J, Price AN, Hajnal JV.
\newblock Complex diffusion-weighted image estimation via matrix recovery under general noise models.
\newblock Neuroimage. 2019 Oct 15;200:391--404.

\bibitem{Akellner2016gibbs}
Kellner E, Dhital B, Kiselev VG, Reisert M.
\newblock Gibbs-ringing artifact removal based on local subvoxel-shifts.
\newblock Magnetic resonance in medicine. 2016;76(5):1574--1581.

\bibitem{Aandersson2003correct}
Andersson JL, Skare S, Ashburner J.
\newblock How to correct susceptibility distortions in spin-echo echo-planar images: application to diffusion tensor imaging.
\newblock Neuroimage. 2003 Oct;20(2):870--888.

\bibitem{Aandersson2016incorporating}
Andersson JL, Graham MS, Zsoldos E, Sotiropoulos SN.
\newblock Incorporating outlier detection and replacement into a non-parametric framework for movement and distortion correction of diffusion MR images.
\newblock Neuroimage. 2016;141:556--572.

\bibitem{Aandersson2016integrated}
Andersson JL, Sotiropoulos SN.
\newblock An integrated approach to correction for off-resonance effects and subject movement in diffusion MR imaging.
\newblock Neuroimage. 2016;125:1063--1078.

\bibitem{Aleemans2009b}
Leemans A, Jones DK.
\newblock The B-matrix must be rotated when correcting for subject motion in DTI data.
\newblock Magnetic Resonance in Medicine: An Official Journal of the International Society for Magnetic Resonance in Medicine. 2009 Jun;61(6):1336--1349.

\bibitem{Azhang2001segmentation}
Zhang Y, Brady M, Smith S.
\newblock Segmentation of brain MR images through a hidden Markov random field model and the expectation-maximization algorithm.
\newblock IEEE transactions on medical imaging. 2001;20(1):45--57.

\bibitem{Asmith2004advances}
Smith SM, Jenkinson M, Woolrich MW, Beckmann CF, Behrens TE, Johansen-Berg H, et~al.
\newblock Advances in functional and structural MR image analysis and implementation as FSL.
\newblock Neuroimage. 2004;23:S208--S219.

\bibitem{ALawhern2018}
Lawhern, V. J., Solon, A. J., Waytowich, N. R., Gordon, S. M., Hung, C. P., Lance, B. J.
\newblock EEGNet: a compact convolutional neural network for EEG-based brain-computer interfaces.
\newblock Journal of neural engineering. 2018;15(5):056013.


\end{thebibliography}


\newpage

\end{document}